# Energy and linear and angular momenta in simple electromagnetic systems

Masud Mansuripur

College of Optical Sciences, The University of Arizona, Tucson, Arizona 85721

**Abstract**. We present examples of simple electromagnetic systems in which energy, linear momentum, and angular momentum exhibit interesting behavior. The systems are sufficiently simple to allow exact solutions of Maxwell's equations in conjunction with the electrodynamic laws of force, torque, energy, and momentum. In all the cases examined, conservation of energy and momentum is confirmed.

**1. Introduction**. This paper presents several examples of simple electromagnetic (EM) systems, whose complete understanding requires careful attention to the detailed behavior of EM force, torque, energy and linear as well as angular momentum. The concept of "hidden momentum" appears in some of the examples, which helps to illustrate the differences between the Einstein-Laub formulation and the standard (i.e., Lorentz) formulation of classical electrodynamics [1-3].

In Sec.2 we analyze the energy and angular momentum content of a finite-duration, finite-diameter wave-packet (or "light bullet") propagating in free space and carrying both spin and orbital angular momentum. Among other findings, we obtain a simple formula for the total angular momentum per photon of the light bullet. The spin contribution to the total angular momentum per photon turns out to be $\hbar\boldsymbol{\sigma}_z$, where $\boldsymbol{\sigma}_z = \boldsymbol{k}_z/k_0 = c\boldsymbol{k}_z/\omega$ is the normalized component of the $k$-vector along $z$ (i.e., the propagation direction), $\omega$ is the angular frequency of the light wave, $c$ is the speed of light in vacuum, and $\hbar$ is the reduced Planck constant.

Section 3 is devoted to an analysis of the energy and angular momentum of a quasi-static system containing electrical charge and current. At first sight, the system appears to violate the law of conservation of angular momentum. We pinpoint the cause of this discrepancy, and point out the whereabouts of the missing angular momentum.

The case of a slowly rotating permanent magnet in an external electric field is taken up in Sec.4. This quasi-static problem reveals the perils and pitfalls of ignoring the contribution of the physical source of the $E$-field (i.e., a pair of electrically-charged parallel plates) when balancing the mechanical momentum imparted to the magnet against the EM momentum contained in the field. Additionally, the same problem, when analyzed in accordance with the Lorentz formalism, shows the important contribution of "hidden momentum" to the physical behavior of the magnet.

In Sec.5 we examine the case of a slowly rotating electric dipole in an external magnetic field. Here, once again, the physical source of the external field (i.e., a solenoid) turns out to be of crucial significance when one attempts to verify the conservation of linear momentum.

Section 6 describes the connection between the energy of a dipole in an external EM field and the energy content of the field in the region surrounding the dipole. Both cases of an electric dipole in an external $E$-field and a magnetic dipole in an external $H$-field (or $B$-field) will be considered. As before, the physical source of the external field will be seen to play an important role when confirming the conservation of energy.

The final section contains a few concluding remarks and general observations. The integrals needed throughout the paper are listed in Appendix A.

**2. A light bullet described by Maxwell's equations in cylindrical coordinates**. Consider a monochromatic wave of frequency $\omega$, propagating in free space along the $z$-axis. The wave-number is $k_z = k_0\sigma_z = (\omega/c)\sigma_z$, where $0 < \sigma_z < 1$. The wave also exhibits an azimuthal phase of $2\pi m$ around the $z$-axis, where $m$, an integer, may be positive, zero, or negative. In cylindrical coordinates, the electric field $\boldsymbol{E}(\boldsymbol{r},t)$ and the magnetic field $\boldsymbol{H}(\boldsymbol{r},t)$ of the wave are given by

$$\boldsymbol{E}(r,\varphi,z,t)=\left[E_r(r)\hat{\boldsymbol{r}}+E_\varphi(r)\hat{\boldsymbol{\varphi}}+E_z(r)\hat{\boldsymbol{z}}\right]\exp[\mathrm{i}(m\varphi+k_z z-\omega t)], \tag{1a}$$

$$\boldsymbol{H}(r,\varphi,z,t)=\left[H_r(r)\hat{\boldsymbol{r}}+H_\varphi(r)\hat{\boldsymbol{\varphi}}+H_z(r)\hat{\boldsymbol{z}}\right]\exp[\mathrm{i}(m\varphi+k_z z-\omega t)]. \tag{1b}$$

Maxwell's equations [4-8] can now be used to relate the various components of the $\boldsymbol{E}$ and $\boldsymbol{H}$ fields to each other, that is,

$$r^{-1}E_r+E_r'+\mathrm{i}mr^{-1}E_\varphi+\mathrm{i}k_zE_z=0, \tag{2}$$

$$mr^{-1}E_z-k_zE_\varphi=\mu_0\omega H_r, \tag{3a}$$

$$k_zE_r+\mathrm{i}E_z'=\mu_0\omega H_\varphi, \tag{3b}$$

$$\mathrm{i}r^{-1}E_\varphi+\mathrm{i}E_\varphi'+mr^{-1}E_r=-\mu_0\omega H_z, \tag{3c}$$

$$mr^{-1}H_z-k_zH_\varphi=-\varepsilon_0\omega E_r, \tag{4a}$$

$$k_zH_r+\mathrm{i}H_z'=-\varepsilon_0\omega E_\varphi, \tag{4b}$$

$$\mathrm{i}r^{-1}H_\varphi+\mathrm{i}H_\varphi'+mr^{-1}H_r=\varepsilon_0\omega E_z, \tag{4c}$$

$$r^{-1}H_r+H_r'+\mathrm{i}mr^{-1}H_\varphi+\mathrm{i}k_zH_z=0. \tag{5}$$

With the help of Eqs.(3a) and (3b), Eq.(4c) may be written as

$$\mathrm{i}r^{-1}(k_zE_r+\mathrm{i}E_z')+\mathrm{i}(k_zE_r'+\mathrm{i}E_z'')+mr^{-1}\left(mr^{-1}E_z-k_zE_\varphi\right)=(\omega/c)^2E_z.$$

Rearranging the terms of the above equation yields

$$\mathrm{i}k_z\left(r^{-1}E_r+E_r'+\mathrm{i}mr^{-1}E_\varphi\right)-E_z''-r^{-1}E_z'+m^2r^{-2}E_z-(\omega/c)^2E_z=0.$$

Invoking Eq.(2), we find

$$E_z''+r^{-1}E_z'+[(\omega/c)^2-k_z^2-(m/r)^2]E_z=0. \tag{6}$$

A similar operation, starting with Eq.(3c) and with the help of Eqs.(4a), (4b) and (5), yields

$$H_z''+r^{-1}H_z'+[(\omega/c)^2-k_z^2-(m/r)^2]H_z=0. \tag{7}$$

We thus find two sets of solutions to Maxwell's equations, one with $E_z=0$, referred to as transverse electric (TE), and another with $H_z=0$, known as transverse magnetic (TM). Defining $k_r=k_0\sigma_r=(\omega/c)\sigma_r$, where $0<\sigma_r<1$ and $\sigma_r^2+\sigma_z^2=1$, the solutions to Eqs.(6) and (7) may be written in terms of Bessel functions of the first kind, order $m$, as

$$E_z(r)=E_{z0}\,J_m(k_rr);\qquad (H_z=0;\ \text{TM}), \tag{8}$$

$$H_z(r)=H_{z0}\,J_m(k_rr);\qquad (E_z=0;\ \text{TE}). \tag{9}$$

In what follows, we will determine the remaining components of the $E$ and $H$ fields for the two cases of TE and TM modes. In each case we also find the energy and angular momentum contained in a finite-diameter, finite duration light pulse.

**2.1. Transverse Electric (TE) modes**: Setting $E_z = 0$ in Eq.(3a), we find $E_\varphi = -(\mu_0\omega/k_z)H_r$. Substitution into Eq.(4b) then yields

$$H_r(r) = \mathrm{i}k_z[(\omega/c)^2 - k_z^2]^{-1}H_z'(r) \quad \rightarrow \quad \begin{cases} H_r(r) = \mathrm{i}(\sigma_z/\sigma_r)H_{z0}\,J_m'(k_r r), \\ E_\varphi(r) = -\mathrm{i}(Z_0/\sigma_r)H_{z0}\,J_m'(k_r r). \end{cases} \tag{8}$$

In the above equation, $Z_0 = \sqrt{\mu_0/\varepsilon_0}$ is the impedance of free space. Subsequently, Eq.(3c) yields

$$E_r(r) = -[mZ_0/(\sigma_r k_r r)]H_{z0}\,J_m(k_r r). \tag{9}$$

Similarly, from Eq.(4a) we find

$$H_\varphi(r) = -[m\sigma_z/(\sigma_r k_r r)]H_{z0}\,J_m(k_r r). \tag{10}$$

The component of the Poynting vector along the $z$-axis is readily obtained as follows:

$$\begin{aligned}\langle S_z(\boldsymbol{r},t)\rangle &= ½\mathrm{Re}\big(E_r H_\varphi^* - E_\varphi H_r^*\big) \\ &= ½[(m^2\sigma_z Z_0)/(k_r\sigma_r r)^2]H_{z0}^2\,J_m^2(k_r r) + ½(Z_0\sigma_z/\sigma_r^2)H_{z0}^2\,{J_m'}^2(k_r r) \\ &= ½(Z_0 H_{z0}^2\sigma_z/\sigma_r^2)\,\big[(m/k_r r)^2\,J_m^2(k_r r) + {J_m'}^2(k_r r)\big].\end{aligned} \tag{11}$$

Integration over the cross-sectional area of the beam from $r = 0$ to $R$ yields

$$\begin{aligned}\int_0^R 2\pi\langle S_z(\boldsymbol{r},t)\rangle r dr &= (\pi Z_0 H_{z0}^2\sigma_z/\sigma_r^2)\int_0^R r\big[(m/k_r r)^2 J_m^2(k_r r) + {J_m'}^2(k_r r)\big]dr \\ &= (\pi Z_0 H_{z0}^2\sigma_z/k_r^2\sigma_r^2)\int_0^{k_r R} x\big[(m/x)^2 J_m^2(x) + {J_m'}^2(x)\big]dx \\ &= (½\pi R^2 Z_0 H_{z0}^2\sigma_z/\sigma_r^2)\{J_{m+1}^2(k_r R) - J_m(k_r R)J_{m+2}(k_r R) + 2m[J_m(k_r R)/(k_r R)]^2\}.\end{aligned} \tag{12}$$

Since Bessel beams in free space travel along the $z$-axis at the speed of $c\sigma_z$, the integrated Poynting vector in Eq.(12) should be equal to the EM energy contained in a cylindrical volume of radius $R$ and length $c\sigma_z$, where $c = 1/\sqrt{\mu_0\varepsilon_0}$ is the speed of light in vacuum. Alternatively, one could calculate the energy content of the cylindrical volume directly, using the total time-averaged energy-density of the $\boldsymbol{E}$ and $\boldsymbol{H}$ fields, as follows:

$$\begin{aligned}\langle \mathcal{E}(\boldsymbol{r},t)\rangle &= ¼\varepsilon_0\mathrm{Re}\big(E_r E_r^* + E_\varphi E_\varphi^*\big) + ¼\mu_0\mathrm{Re}\big(H_r H_r^* + H_\varphi H_\varphi^* + H_z H_z^*\big) \\ &= ¼\varepsilon_0[mZ_0/(\sigma_r k_r r)]^2 H_{z0}^2\,J_m^2(k_r r) + ¼\varepsilon_0(Z_0/\sigma_r)^2 H_{z0}^2\,{J_m'}^2(k_r r) \\ &\quad + ¼\mu_0(\sigma_z/\sigma_r)^2 H_{z0}^2\,{J_m'}^2(k_r r) + ¼\mu_0[m\sigma_z/(\sigma_r k_r r)]^2 H_{z0}^2\,J_m^2(k_r r) + ¼\mu_0 H_{z0}^2\,J_m^2(k_r r) \\ &= ¼\mu_0[2m^2/(\sigma_r k_r r)^2 - (m/k_r r)^2 + 1]H_{z0}^2\,J_m^2(k_r r) + ¼\mu_0[(2/\sigma_r^2) - 1]H_{z0}^2\,{J_m'}^2(k_r r) \\ &= ½(\mu_0 H_{z0}^2/\sigma_r^2)\big[(m/k_r r)^2\,J_m^2(k_r r) + {J_m'}^2(k_r r)\big] \\ &\quad + ¼\mu_0 H_{z0}^2\,\big[J_m^2(k_r r) - (m/k_r r)^2\,J_m^2(k_r r) - {J_m'}^2(k_r r)\big].\end{aligned} \tag{13}$$

In unit time, the Bessel beam propagates a distance $c\sigma_z$ along the $z$-axis. Multiplying $c\sigma_z$ into Eq.(13) yields a first term that is identical to Eq.(11). The second term, therefore, must integrate to zero over the cross-sectional area of the beam. The integral of the second term on the right-hand-side of Eq.(13) over the cylindrical volume of radius $R$ and length $c\sigma_z$ may be written

$$\tfrac{1}{2}\pi\sigma_z Z_0 H_{z0}^2 \int_0^R \{J_m^2(k_r r) - [(m/k_r r)\, J_m(k_r r) - J_m'(k_r r)]^2 - 2(m/k_r r)\, J_m(k_r r)\, J_m'(k_r r)\}\, r dr$$

$$= \tfrac{1}{2}\pi(\sigma_z/k_r^2) Z_0 H_{z0}^2 \int_0^{k_r R} [x J_m^2(x) - x J_{m+1}^2(x) - 2m\, J_m(x)\, J_m'(x)]\, dx$$

$$= \tfrac{1}{2}\pi(\sigma_z/k_r^2) Z_0 H_{z0}^2 [k_r R\, J_{m+1}(k_r R) - m J_m(k_r R)]\, J_m(k_r R). \qquad (14)$$

The above expression vanishes when $R$ is chosen to correspond to one of the zeroes of $J_m(k_r R)$. The energy content of a cylindrical volume of radius $R$ and length $c\sigma_z$ is thus given by Eq.(12) provided that $R$ is chosen such that $J_m(k_r R) = 0$.

Next we compute the angular momentum density of the TE mode, as follows:

$$\langle L_z(\boldsymbol{r}, t)\rangle \hat{\boldsymbol{z}} = r\hat{\boldsymbol{r}} \times \tfrac{1}{2}\mathrm{Re}(-E_r H_z^* \hat{\boldsymbol{\varphi}}/c^2) = \tfrac{1}{2}[m Z_0 H_{z0}^2/(c^2 k_r \sigma_r)]\, J_m^2(k_r r)\hat{\boldsymbol{z}}. \qquad (15)$$

Integrating the above expression over the cross-sectional area of the beam from $r = 0$ to $R$ yields

$$\int_0^R 2\pi r \langle L_z(\boldsymbol{r}, t)\rangle dr = [\pi m Z_0 H_{z0}^2/(c^2 k_r \sigma_r)] \int_0^R r\, J_m^2(k_r r) dr$$

$$= [\pi m Z_0 H_{z0}^2/(c^2 k_r^3 \sigma_r)] \int_0^{k_r R} x\, J_m^2(x) dx$$

$$= [\tfrac{1}{2} m(\pi R^2 Z_0 H_{z0}^2)/(c^2 k_r \sigma_r)][J_m^2(k_r R) - J_{m-1}(k_r R)\, J_{m+1}(k_r R)]$$

$$= [\tfrac{1}{2} m(\pi R^2 Z_0 H_{z0}^2)/(c\omega \sigma_r^2)][J_m^2(k_r R) - J_{m-1}(k_r R)\, J_{m+1}(k_r R)]. \qquad (16)$$

In unit time the beam propagates a distance of $c\sigma_z$, so the angular momentum in Eq.(16) must be multiplied by $c\sigma_z$ in order to correspond to the same amount of energy that is represented by Eq.(12). The ratio of angular momentum to energy content of a cylindrical volume of radius $R$ and arbitrary length, assuming that $J_m(k_r R) = 0$, is thus given by $-(m/\omega) J_{m-1}(k_r R)/J_{m+1}(k_r R)$. For sufficiently large $R$, the asymptotic value of $J_m(k_r R)$ is $\sqrt{2/(\pi k_r R)}\cos(k_r R - \tfrac{1}{2}m\pi - \tfrac{1}{4}\pi)$, which leads to $J_{m-1}(k_r R)/J_{m+1}(k_r R) \cong -1$. Consequently, the ratio of angular momentum to energy for a sufficiently large number of rings of the TE mode of a Bessel beam in vacuum is given by $m/\omega$. In quantum-optical terms, where the energy of each photon is $\hbar\omega$, the angular momentum per photon will be $m\hbar$.

**2.2. Transverse Magnetic (TM) modes**: Setting $H_z = 0$ in Eq.(4a), we find $H_\varphi = (\varepsilon_0 \omega/k_z) E_r$. Substitution into Eq.(3b) then yields

$$E_r(r) = \mathrm{i} k_z [(\omega/c)^2 - k_z^2]^{-1} E_z'(r) \quad \rightarrow \quad \begin{cases} E_r(r) = \mathrm{i}(\sigma_z/\sigma_r) E_{z0}\, J_m'(k_r r), \\ H_\varphi(r) = \mathrm{i}(E_{z0}/Z_0 \sigma_r)\, J_m'(k_r r). \end{cases} \qquad (17)$$

Subsequently, Eq.(4c) yields

$$H_r(r) = (m E_{z0}/Z_0 k_r \sigma_r r) J_m(k_r r). \qquad (18)$$

Similarly, from Eq.(3a) we find

$$E_\varphi(r) = -(m\sigma_z/\sigma_r k_r r) E_{z0}\, J_m(k_r r). \qquad (19)$$

Having determined the various components of the $\boldsymbol{E}$ and $\boldsymbol{H}$ fields of the TM mode of a Bessel beam in vacuum, we proceed to compute the Poynting vector along the $z$-axis, as follows:

$$\langle S_z(\boldsymbol{r},t)\rangle = ½\text{Re}\big(E_r H_\varphi^* - E_\varphi H_r^*\big)$$
$$= ½[(\sigma_z E_{z0}^2)/(Z_0\sigma_r^2)]\, {J'_m}^2(k_r r) + ½(\sigma_z E_{z0}^2/Z_0)[m/(\sigma_r k_r r)]^2\, J_m^2(k_r r)$$
$$= ½[(\sigma_z E_{z0}^2)/(Z_0\sigma_r^2)]\, \big[{J'_m}^2(k_r r) + (m/k_r r)^2\, J_m^2(k_r r)\big]. \qquad (20)$$

Integration over the cross-sectional area of the beam from $r = 0$ to $R$ yields

$$\int_0^R 2\pi r\langle S_z(\boldsymbol{r},t)\rangle dr = ½(\pi R^2\sigma_z E_{z0}^2/Z_0\sigma_r^2)$$
$$\times\{J_{m+1}^2(k_r R) - J_m(k_r R)J_{m+2}(k_r R) + 2m[J_m(k_r R)/(k_r R)]^2\}. \qquad (21)$$

Aside from the coefficient $E_{z0}^2/Z_0$, this result is identical to that obtained in Eq.(12) for a TE beam, where the corresponding coefficient was $Z_0 H_{z0}^2$.

In similar fashion, the angular momentum density of a TM Bessel beam is found to be

$$\langle L_z(\boldsymbol{r},t)\rangle\hat{\boldsymbol{z}} = r\hat{\boldsymbol{r}} \times ½\text{Re}(E_z H_r^*\hat{\boldsymbol{\varphi}}/c^2) = ½[mE_{z0}^2/(Z_0 c^2 k_r\sigma_r)]\, J_m^2(k_r r)\hat{\boldsymbol{z}}. \qquad (22)$$

Integration over the cross-sectional area of the beam from $r = 0$ to $R$ yields

$$\int_0^R 2\pi r\langle L_z(\boldsymbol{r},t)\rangle dr = [\pi m E_{z0}^2/(Z_0 c^2 k_r\sigma_r)]\int_0^R r\, J_m^2(k_r r)dr$$
$$= [\pi m E_{z0}^2/(Z_0 c^2 k_r^3\sigma_r)]\int_0^{k_r R} x\, J_m^2(x)dx$$
$$= ½m[(\pi R^2 E_{z0}^2)/(Z_0 c^2 k_r\sigma_r)][J_m^2(k_r R) - J_{m-1}(k_r R)\, J_{m+1}(k_r R)]$$
$$= ½m[(\pi R^2 E_{z0}^2)/(Z_0 c\omega\sigma_r^2)][J_m^2(k_r R) - J_{m-1}(k_r R)\, J_{m+1}(k_r R)]. \qquad (23)$$

In unit time the beam propagates a distance of $c\sigma_z$, so the angular momentum in Eq.(23) must be multiplied by $c\sigma_z$ in order to correspond to the same amount of energy that is represented by Eq.(21). Once again, the ratio of angular momentum to energy contained in a cylindrical volume of radius $R$ and arbitrary length, where $J_m(k_r R) = 0$, is seen to approach $m/\omega$ in the limit of large $R$. In quantum-optical terms, each photon of energy $\hbar\omega$ in the TM mode of the Bessel beam carries $m\hbar$ of orbital angular momentum.

**2.3. The case of circular polarization**. We *define* a circularly-polarized Bessel beam as an equal mixture of TE and TM modes, with the phase difference between the two eigen-modes being $\pm 90°$ for right/left circular states. Thus, in the $k$-space, all plane-waves that constitute the resulting Bessel beam will be circularly polarized. For a pair of TE and TM beams having the same $\omega$, $k_z$, and $m$, and also having $E_{z0} = \pm\text{i}Z_0 H_{z0}$, the superposed $\boldsymbol{E}$ and $\boldsymbol{H}$ fields are given by

$$E_r(r) = \pm\text{i}[mE_{z0}/(\sigma_r k_r r)]\, J_m(k_r r) + \text{i}(\sigma_z/\sigma_r)E_{z0}\, J'_m(k_r r). \qquad (24)$$

$$E_\varphi(r) = \mp(E_{z0}/\sigma_r)\, J'_m(k_r r) - (m\sigma_z/\sigma_r k_r r)E_{z0}\, J_m(k_r r). \qquad (25)$$

$$E_z(r) = E_{z0}\, J_m(k_r r). \qquad (26)$$

$$H_r(r) = \pm(E_{z0}\sigma_z/Z_0\sigma_r)\, J'_m(k_r r) + (mE_{z0}/Z_0 k_r\sigma_r r)J_m(k_r r). \qquad (27)$$

$$H_\varphi(r) = \pm\text{i}[m\sigma_z E_{z0}/(Z_0\sigma_r k_r r)]\, J_m(k_r r) + \text{i}(E_{z0}/Z_0\sigma_r)\, J'_m(k_r r). \qquad (28)$$

$$H_z(r) = \mp\text{i}(E_{z0}/Z_0)\, J_m(k_r r). \qquad (29)$$

The angular momentum density of this circularly-polarized Bessel beam is readily found to be

$$\langle L_z(\boldsymbol{r},t)\rangle\hat{\boldsymbol{z}} = r\hat{\boldsymbol{r}} \times ½\text{Re}[(E_z H_r^* - E_r H_z^*)\hat{\boldsymbol{\varphi}}/c^2]$$

$$= [E_{z0}^2/(Z_0 c^2 \sigma_r k_r)][mJ_m^2(k_r r) \pm \sigma_z k_r r J_m(k_r r) J_m'(k_r r)]\hat{\boldsymbol{z}}. \qquad (30)$$

Upon integration over the cross-sectional area of the beam (from $r = 0$ to $R$) we find

$$\int_0^R 2\pi r\langle L_z(\boldsymbol{r},t)\rangle dr =$$

$$= [2\pi E_{z0}^2/(Z_0 c^2 \sigma_r k_r)] \int_0^R [mJ_m^2(k_r r) \pm \sigma_z k_r r J_m(k_r r) J_m'(k_r r)] r dr$$

$$= [2\pi E_{z0}^2/(Z_0 c^2 \sigma_r k_r^3)] \int_0^{k_r R} [mxJ_m^2(x) \pm \sigma_z x^2 J_m(x) J_m'(x)] dx$$

$$= [2\pi E_{z0}^2/(Z_0 c^2 \sigma_r k_r^3)]$$

$$\times \{½m(k_r R)^2[J_m^2(k_r R) - J_{m-1}(k_r R)J_{m+1}(k_r R)] \pm ½\sigma_z (k_r R)^2 J_{m-1}(k_r R) J_{m+1}(k_r R)\}$$

$$= [\pi R^2 E_{z0}^2/(Z_0 c\omega\sigma_r^2)][(\pm\sigma_z - m)J_{m-1}(k_r R)J_{m+1}(k_r R) + mJ_m^2(k_r R)]. \qquad (31)$$

If $R$ is chosen to correspond to a zero of $J_m(k_r R)$, the last term in Eq.(31) vanishes, and the remaining terms give the angular momentum due to both vorticity and circular polarization.

Finally, we compute the $z$-component of the Poynting vector of the circularly-polarized Bessel beam, as follows:

$$\langle S_z(\boldsymbol{r},t)\rangle = ½\text{Re}\big(E_r H_\varphi^* - E_\varphi H_r^*\big)$$

$$= ½\{[m^2\sigma_z E_{z0}^2/Z_0(\sigma_r k_r r)^2]\, J_m^2(k_r r) \pm [mE_{z0}^2/(Z_0\sigma_r^2 k_r r)]\, J_m(k_r r)\, J_m'(k_r r)$$

$$\pm[m\sigma_z^2 E_{z0}^2/(Z_0\sigma_r^2 k_r r)]\, J_m(k_r r) J_m'(k_r r) + (\sigma_z E_{z0}^2/Z_0\sigma_r^2) {J_m'}^2(k_r r)$$

$$+ (\sigma_z E_{z0}^2/Z_0\sigma_r^2)\, {J_m'}^2(k_r r) \pm (mE_{z0}^2/Z_0 k_r \sigma_r^2 r) J_m(k_r r) J_m'(k_r r)$$

$$\pm(m\sigma_z^2 E_{z0}^2/Z_0\sigma_r^2 k_r r) J_m(k_r r) J_m'(k_r r) + [m^2\sigma_z E_{z0}^2/Z_0(\sigma_r k_r r)^2]\, J_m^2(k_r r)\}$$

$$= (\sigma_z E_{z0}^2/Z_0\sigma_r^2)\big[(m/k_r r)^2 J_m^2(k_r r) + {J_m'}^2(k_r r)\big]$$

$$\pm[m(1+\sigma_z^2)E_{z0}^2/(Z_0\sigma_r^2 k_r r)]\, J_m(k_r r) J_m'(k_r r). \qquad (32)$$

The first term on the right-hand-side of the above equation is twice the expression in Eq.(20). Therefore, upon integration, this term yields the total power of the beam, as before. The second term integrates to

$$\pm[2\pi m(1+\sigma_z^2)E_{z0}^2/(Z_0\sigma_r^2 k_r^2)] \int_0^{k_r R} J_m(x) J_m'(x) dx = \pm[\pi m(1+\sigma_z^2)E_{z0}^2/(Z_0\sigma_r^2 k_r^2)]\, J_m^2(k_r R). \quad (33)$$

This function vanishes if $R$ is chosen to coincide with a zero of $J_m(k_r R)$. Thus the energy content of our circularly-polarized Bessel beam is equal to the sum of the energies of its TE and TM components. The ratio of angular momentum to energy (in the limit of large $R$) is now seen to be $(m \mp \sigma_z)/\omega$, which includes the contribution $\pm\sigma_z/\omega$ of spin to the orbital angular momentum $m/\omega$. The $\pm$ sign is related to the sense of circular polarization, which could be the same as or opposite to the sense of the orbital angular momentum of the Bessel beam of order $m$. When the Bessel beam has a large diameter (compared to a wavelength $\lambda = 2\pi c/\omega$), its $\sigma_z$ will be close to 1.0, in which case the spin angular momentum per photon will be nearly $\pm\hbar$. For more compact Bessel beams, however, the spin angular momentum per photon will be $\pm\hbar\sigma_z$.

**3. Electromagnetic energy and angular momentum of a rotating cylinder**. As a second example of simple EM systems, consider an infinitely long, uniformly-charged, non-conducting, hollow cylinder of radius $R$, having a surface-charge-density $\sigma_s$, as shown in Fig.1(a). The $E$-field inside the cylinder is zero, while that outside may be evaluated using Gauss's law [4,5], as follows:

$$\boldsymbol{E}_0(\boldsymbol{r}) = R\sigma_s\hat{\boldsymbol{r}}/(\varepsilon_0 r)\,; \qquad r > R. \qquad (34)$$

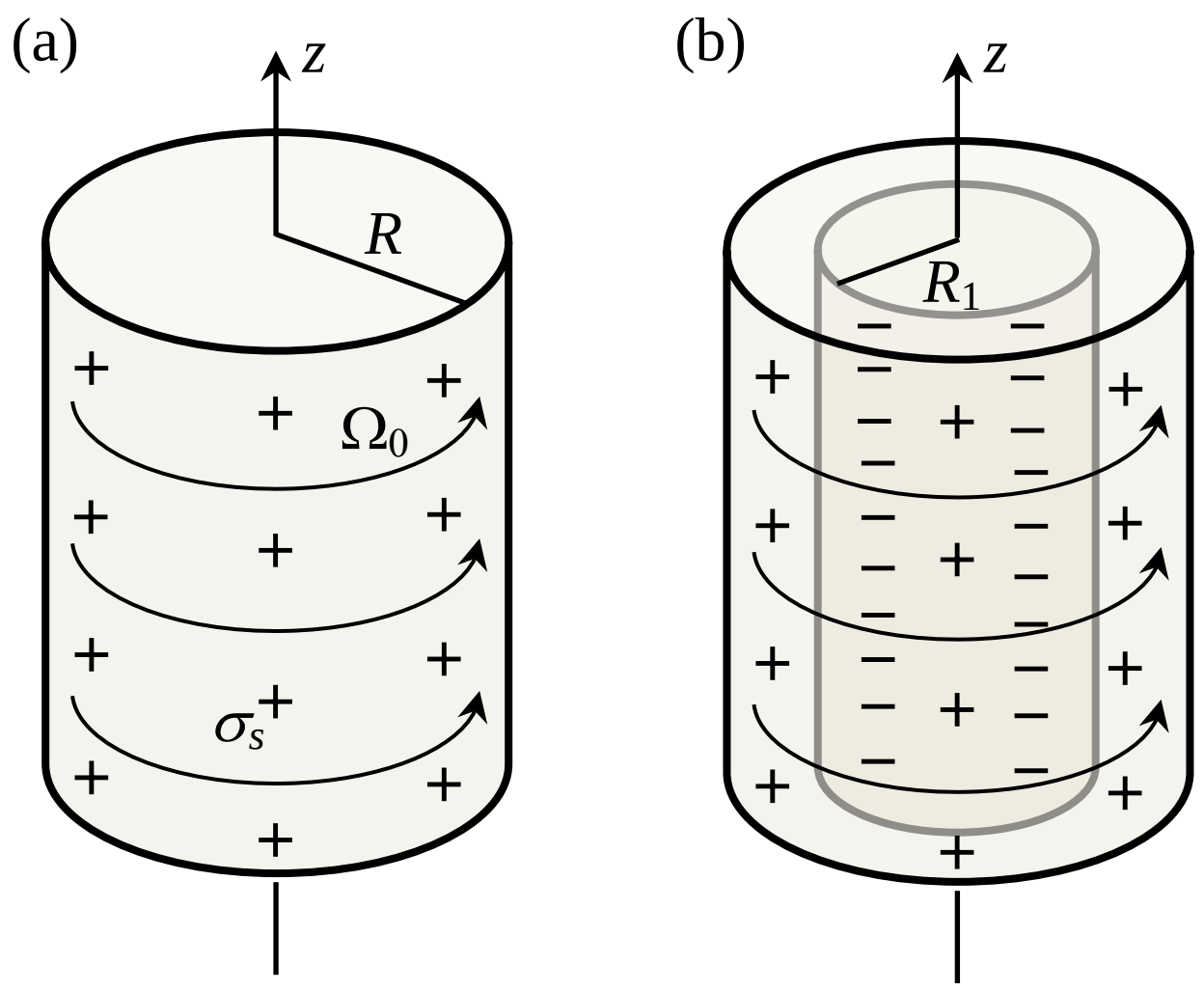

**Fig.1**. (a) Infinitely-long hollow cylinder of radius $R$, having a uniform surface-charge-density $\sigma_s$, is brought from rest to a constant angular velocity $\Omega_0$ around the $z$-axis. (b) A second infinitely-long cylinder of radius $R_1$ and uniform surface-charge-density $\sigma_{s1}$ resides inside the cylinder of radius $R$. The inner cylinder, which does *not* rotate, is concentric with the outer (rotating) cylinder.

Let the cylinder start to spin slowly around its axis, so that its angular velocity is brought from zero to some finite value $\Omega_0$ after a long time, $t_{\max}$. During the time interval $(0, t_{\max})$, when the angular velocity is $\Omega(t)$, the surface-current-density will be $\boldsymbol{J}_s(t) = \sigma_s R\Omega(t)\hat{\boldsymbol{\varphi}}$. Ignoring the retardation effects (justified by the gradual rise of the spinning velocity), the vector potential $\boldsymbol{A}(\boldsymbol{r}, t)$ may be calculated as follows [4-8]:

$$\boldsymbol{A}(\boldsymbol{r},t) = \frac{\mu_0}{4\pi}\iint \frac{\boldsymbol{J}_s(\boldsymbol{r}',t)}{|\boldsymbol{r}-\boldsymbol{r}'|}ds' = \frac{\mu_0}{4\pi}\int_{\varphi=0}^{\pi}\int_{z=-\infty}^{\infty}\frac{2\sigma_s R\Omega\cos\varphi}{\sqrt{(r^2+R^2-2rR\cos\varphi)+z^2}}\hat{\boldsymbol{\varphi}}Rd\varphi dz$$

$$= \frac{\mu_0\sigma_s R^2\Omega\,\hat{\boldsymbol{\varphi}}}{2\pi}\int_0^{\pi}\cos\varphi\lim_{z\to\infty}\left\{\ln\frac{\sqrt{(r^2+R^2-2rR\cos\varphi)+z^2}+z}{\sqrt{(r^2+R^2-2rR\cos\varphi)+z^2}-z}\right\}d\varphi$$ ← See Appendix A

$$= \frac{\mu_0\sigma_s R^2\Omega\hat{\boldsymbol{\varphi}}}{2\pi}\int_0^{\pi}\cos\varphi\lim_{z\to\infty}\left\{\ln\frac{\left[\sqrt{(r^2+R^2-2rR\cos\varphi)+z^2}+z\right]^2}{r^2+R^2-2rR\cos\varphi}\right\}d\varphi$$

$$= \frac{\mu_0\sigma_s R^2\Omega\hat{\boldsymbol{\varphi}}}{2\pi}\int_0^{\pi}\cos\varphi\lim_{z\to\infty}\left\{2\ln\left[\sqrt{(r^2+R^2-2rR\cos\varphi)+z^2}+z\right]\right\}d\varphi$$

$$-\frac{\mu_0\sigma_s R^2\Omega\hat{\boldsymbol{\varphi}}}{2\pi}\int_0^{\pi}\cos\varphi\ln(r^2+R^2-2rR\cos\varphi)\,d\varphi$$

$$= \frac{\mu_0\sigma_s R^2\Omega\hat{\boldsymbol{\varphi}}}{\pi}\int_0^{\pi}\cos\varphi\lim_{z\to\infty}\left\{\ln(z)+\ln\left[1+\sqrt{1+(r^2+R^2-2rR\cos\varphi)/z^2}\right]\right\}d\varphi$$

$$-\frac{\mu_0\sigma_s R^2\Omega\hat{\boldsymbol{\varphi}}}{2\pi}\int_0^{\pi}\cos\varphi\,\{\ln(R^2)+\ln[1-2(r/R)\cos\varphi+(r/R)^2]\}d\varphi$$

$$= \frac{\mu_0\sigma_s R^2\Omega\hat{\boldsymbol{\varphi}}}{\pi}[\lim_{z\to\infty}\ln(z)+\ln(2)-\ln(R)]\int_0^{\pi}\cos\varphi\,d\varphi$$

$$-\frac{\mu_0\sigma_s R^2\Omega\hat{\boldsymbol{\varphi}}}{2\pi}\int_0^{\pi}\cos\varphi\ln[1-2(r/R)\cos\varphi+(r/R)^2]\,d\varphi$$ ← See Appendix A

$$= \begin{cases} ½(\mu_0\sigma_s R\Omega r)\hat{\boldsymbol{\varphi}}, & r < R; \\ ½(\mu_0\sigma_s R^3\Omega/r)\hat{\boldsymbol{\varphi}}, & r > R. \end{cases} \quad (35)$$

The magnetic field produced by the spinning cylinder is thus found to be

$$\boldsymbol{B}(\boldsymbol{r},t) = \mu_0\boldsymbol{H}(\boldsymbol{r},t) = \boldsymbol{\nabla}\times\boldsymbol{A}(\boldsymbol{r},t) = r^{-1}\partial_r(rA_\varphi)\hat{\boldsymbol{z}} = \begin{cases} \mu_0\sigma_s R\Omega(t)\hat{\boldsymbol{z}} & r < R, \\ 0 & r > R. \end{cases} \quad (36)$$

While the cylinder spins up, an electric field is induced by the time-varying vector potential. The magnitude of this $E$-field at the cylinder surface is

$$\boldsymbol{E}(r = R,t) = -\partial_t\boldsymbol{A}(r = R,t) = -½(\mu_0\sigma_s R^2\Omega')\hat{\boldsymbol{\varphi}}. \quad (37)$$

The mechanical energy (per unit length along the $z$-axis) needed to bring the cylinder from rest to its final spinning speed $\Omega_0 = \Omega(t_{\max})$ is obtained by integrating $\boldsymbol{E}\cdot\boldsymbol{J}_s$ over the cylindrical surface and over time from $t = 0$ to $t_{\max}$. We will have

$$\begin{aligned} \mathcal{E} &= 2\pi R\int_0^{t_{\max}} E(r = R,t)\sigma_s R\Omega(t)dt \\ &= \mu_0\pi R^4\sigma_s^2\int_0^{t_{\max}}\Omega'(t)\Omega(t)dt = ½\mu_0\pi R^4\sigma_s^2\Omega_0^2. \end{aligned} \quad (38)$$

This energy is stored in the internal $B$-field (or $H$-field) of the cylinder, whose density is $½\mu_0 H^2 = ½\mu_0(\sigma_s R\Omega_0)^2$.

**3.1. Angular momentum conservation**. In the system of Fig.1(a), the torque (per unit length along the $z$-axis) exerted on the cylinder is obtained by integrating $\boldsymbol{r}\times\sigma_s\boldsymbol{E}(r = R,t)$ over the cylindrical surface, that is,

$$\boldsymbol{T}(t) = \int_{\varphi=0}^{2\pi} R\hat{\boldsymbol{r}}\times(-½\mu_0\sigma_s^2 R^2\Omega')\hat{\boldsymbol{\varphi}}Rd\varphi = -\pi\mu_0\sigma_s^2 R^4\Omega'(t)\hat{\boldsymbol{z}}. \quad (39)$$

The total angular momentum picked up by the agency in charge of spinning the cylinder is thus given by

$$\boldsymbol{L} = \int_0^{t_{\max}}\boldsymbol{T}(t)dt = -\pi\mu_0\sigma_s^2 R^4\Omega_0\hat{\boldsymbol{z}}. \quad (40)$$

At this point, we are faced with a dilemma. In its final state, the charged cylinder is spinning at the constant angular velocity $\Omega_0$, producing a constant, uniform magnetic field $\boldsymbol{H} = \sigma_s R\Omega_0\hat{\boldsymbol{z}}$ inside the cylinder, and a constant, radially-decaying electric field $\boldsymbol{E}_0(r)$ given by Eq.(34) outside the cylinder. Therefore, the EM field produced by the spinning charged cylinder appears to *not* have any angular momentum. In other words, the existence of the mechanical angular momentum $\boldsymbol{L}$ of Eq.(40), which is acquired by the spinning agent, appears to contradict the conservation of angular momentum. This discrepancy, however, is easily resolved if one recognizes the assumed infinite length of the cylinder as the culprit. A finite-length cylinder will always produce a magnetic field in faraway regions ($r \gg R$), and the EM angular momentum density $\boldsymbol{\mathcal{L}}_{em}(\boldsymbol{r}) = \boldsymbol{r}\times(\boldsymbol{E}\times\boldsymbol{H}/c^2)$, when integrated over the entire space, turns out to be exactly equal in magnitude and opposite in direction to the mechanical angular momentum $\boldsymbol{L}$ of Eq.(40).

A more tractable version of the above problem involves a second charged cylinder of radius $R_1 < R$ placed inside (and concentric with) the spinning cylinder, as shown in Fig.1(b). If the surface-charge-density of the inner cylinder is taken to be $\sigma_{s1} = -R\sigma_s/R_1$, the static $E$-field will be confined to the region between the two cylinders, as follows:

$$\boldsymbol{E}_0(\boldsymbol{r}) = \begin{cases} 0, & r < R_1, \\ -R\sigma_s\hat{\boldsymbol{r}}/(\varepsilon_0 r), & R_1 < r < R, \\ 0, & r > R. \end{cases} \tag{41}$$

The EM angular momentum (per unit length along $z$), which is now confined to the region between the two cylinders, is given by

$$\boldsymbol{L}_{em} = \int_{\varphi=0}^{2\pi}\int_{r=R_1}^{R} \boldsymbol{r} \times (\boldsymbol{E}_0 \times \boldsymbol{H}/c^2) r dr d\varphi = \pi\mu_0\sigma_s^2 R^2(R^2 - R_1^2)\Omega_0\hat{\boldsymbol{z}}. \tag{42}$$

The difference between the mechanical angular momentum $\boldsymbol{L}$ in Eq.(40) and the EM angular momentum in Eq.(42) is taken up by the external agency responsible for holding the inner cylinder stationary while the outer cylinder is spun from rest to its final angular velocity $\Omega_0$. Needless to say, no energy is spent in keeping the inner cylinder stationary. However, the induced $E$-field $\boldsymbol{E}(r = R_1, t) = -\partial_t \boldsymbol{A}(r = R_1, t) = -½(\mu_0\sigma_s R R_1 \Omega')\hat{\boldsymbol{\varphi}}$ exerts a torque on the inner cylinder, which, upon integration over time, yields

$$\boldsymbol{L}_1 = \int_{t=0}^{t_{\max}}\int_{\varphi=0}^{2\pi} R_1\hat{\boldsymbol{r}} \times \sigma_{s1}\boldsymbol{E}(r = R_1, t) R_1 d\varphi dt = \pi\mu_0\sigma_s^2 R^2 R_1^2 \Omega_0\hat{\boldsymbol{z}}. \tag{43}$$

The total mechanical angular momentum $\boldsymbol{L} + \boldsymbol{L}_1$ is thus seen to be exactly equal in magnitude and opposite in direction to the EM angular momentum $\boldsymbol{L}_{em}$ stored in the $\boldsymbol{E}$ and $\boldsymbol{H}$ fields in the region between the two cylinders.

A similar argument can be made if the oppositely-charged stationary cylinder has a radius $R_1 > R$. The EM momentum now vanishes, and the angular momenta of the cylinders cancel out.

**3.2. Spinning cylinder in an external electric field**. Next, we assume that the rotating cylinder of Fig.1(a) is placed inside a uniform electric field $\boldsymbol{E} = E_x\hat{\boldsymbol{x}}$, as shown in Fig.2. The charges on the cylinder and those on the plates attract/repel each other along the $x$-axis. However, no other net force is exerted on the cylinder as it spins up from $\Omega = 0$ to $\Omega_0$. In contrast, the induced $E$-field during the time-interval $(0, t_{\max})$, namely,

$$\boldsymbol{E}(r, t) = -\partial_t \boldsymbol{A}(r, t) = -½(\mu_0\sigma_s R^3 \Omega'/r)\hat{\boldsymbol{\varphi}}, \tag{44}$$

exerts a force in the $y$-direction on both plates. With reference to Fig.2 and considering that $y = (d/2)\tan\varphi$, the force (per unit-length along $z$) exerted on each plate is readily seen to be

$$F_y(t) = \int_{\varphi=-½\pi}^{½\pi} -\sigma_0 E_\varphi\left(r = \frac{d}{2\cos\varphi}, t\right)\cos\varphi \underbrace{(d/2)(1 + \tan^2\varphi)d\varphi}_{dy} = ½\pi\mu_0\sigma_0\sigma_s R^3\Omega'. \tag{45}$$

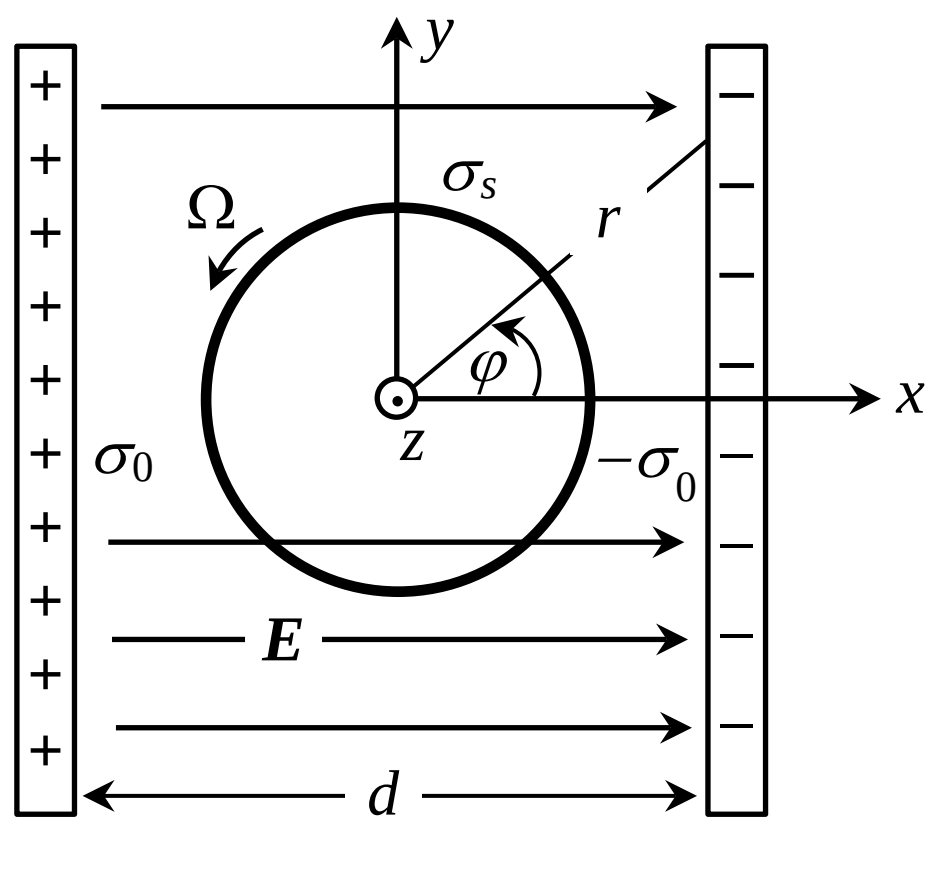

**Fig.2**. An infinitely-long, uniformly-charged cylinder of radius $R$ and surface-charge-density $\sigma_s$ rotates at an angular velocity $\Omega$ around the $z$-axis. A pair of uniformly-charged parallel plates produces a constant, uniform $E$-field in the cross-sectional plane of the cylinder. The surface-charge-density of the parallel plates is $\pm\sigma_0$, the $E$-field produced in the space between the plates is $\boldsymbol{E} = (\sigma_0/\varepsilon_0)\hat{\boldsymbol{x}}$, the $H$-field inside the cylinder is $\boldsymbol{H} = \sigma_s R\Omega\,\hat{\boldsymbol{z}}$, and the EM momentum-density within the cylinder is given by $\boldsymbol{\wp}_{em} = \boldsymbol{E} \times \boldsymbol{H}/c^2 = -(\mu_0\sigma_0\sigma_s R\Omega)\,\hat{\boldsymbol{y}}$.

When the above force is integrated over time from $t = 0$ to $t_{\max}$ and multiplied by two (to account for both plates), we find the mechanical momentum acquired by the plates at $t = t_{\max}$ to be $\boldsymbol{p} = \pi\mu_0\sigma_0\sigma_s R^3\Omega_0\hat{\boldsymbol{y}}$. Considering that the $E$-field produced by the plates is $\boldsymbol{E} = (\sigma_0/\varepsilon_0)\hat{\boldsymbol{x}}$, and the $H$-field inside the cylinder is $\boldsymbol{H} = \sigma_s R\Omega_0\hat{\boldsymbol{z}}$, the EM momentum-density within the cylinder will be $\boldsymbol{\wp}_{em} = \boldsymbol{E}\times\boldsymbol{H}/c^2 = -(\mu_0\sigma_0\sigma_s R\Omega_0)\hat{\boldsymbol{y}}$. Multiplication by the cross-sectional area of the cylinder then shows that the EM momentum stored within the cylinder is precisely equal in magnitude and opposite in sign to the mechanical momentum acquired by the plates.

**3.3. Hidden momentum of the spinning cylinder**. It has been argued that a stationary system such as that of the spinning cylinder residing between the parallel plates of Fig.2 must have an internal "hidden momentum" equal and opposite to its EM momentum, so that the net linear momentum of the stationary system would be zero [9-32]. The acquisition of hidden momentum by the cylinder thus dictates the transfer of a compensatory mechanical momentum to the cylinder as it spins up from rest ($\Omega = 0$) to its final state where $\Omega = \Omega_0$. The linear mechanical momentum per unit-length of the cylinder at $t = t_{\max}$ is thus argued to be $\boldsymbol{p}_{\text{mech}} = -\pi\mu_0\sigma_0\sigma_s R^3\Omega_0\hat{\boldsymbol{y}}$. In this way, the system acquires no net mechanical momentum, as the cylinder and the plates end up moving in opposite directions with equal momentum, while the EM momentum is compensated by an equal and opposite amount of hidden momentum.

If the spinning cylinder happens to be used as a model for a uniformly-magnetized solid cylinder, then two alternative formulations of classical electrodynamics may be used to describe the situation discussed above. In the Lorentz formalism [4,5], the magnetization $M_0\hat{\boldsymbol{z}}$ of the cylinder is equivalent to a current-density $\boldsymbol{J} = \mu_0^{-1}\boldsymbol{\nabla}\times\boldsymbol{M}$, which turns out to be the surface current-density $\boldsymbol{J}_s = \mu_0^{-1}M_0\hat{\boldsymbol{\varphi}}$. The $H$-field inside the cylinder is now zero, but the internal $B$-field continues to be given by $\boldsymbol{B} = \mu_0 J_s\hat{\boldsymbol{z}}$, which is equal to $M_0\hat{\boldsymbol{z}}$. In the Lorentz formalism, the EM momentum-density is $\boldsymbol{\wp}_{em} = \varepsilon_0\boldsymbol{E}\times\boldsymbol{B}$ (known as Livens momentum), whereas the hidden momentum-density is $\boldsymbol{\wp}_{\text{hidden}} = \boldsymbol{M}\times\varepsilon_0\boldsymbol{E}$. Therefore, all the preceding arguments remain intact.

In the Einstein-Laub formalism [33], where the EM momentum-density is $\boldsymbol{\wp}_{em} = \boldsymbol{E}\times\boldsymbol{H}/c^2$, there is no need for hidden momentum. In this case, the EM momentum of the system is zero. However, the force-density exerted by an external $E$-field on the magnetization $\boldsymbol{M}(\boldsymbol{r},t)$ is given by $\boldsymbol{f}(\boldsymbol{r},t) = -\partial_t\boldsymbol{M}\times\varepsilon_0\boldsymbol{E}$. If $\boldsymbol{M}$ rises slowly from zero to $M_0\hat{\boldsymbol{z}}$, the mechanical momentum-density acquired by the magnetized cylinder will be $-\varepsilon_0 M_0 E_x\hat{\boldsymbol{y}}$, which is the same mechanical momentum as obtained in the Lorentz formalism — albeit after invoking the notion of hidden momentum, an endemic feature of the Lorentz formulation in its treatment of magnetic materials.

**4. Rotating magnetic dipole in a constant and uniform electric field**. As our third example, we consider a small magnetic dipole $\boldsymbol{m}_0$ residing at the origin of the $xy$-plane and rotating slowly around the $z$-axis at the constant angular velocity $\omega$, as shown in Fig. 3. We have

$$\boldsymbol{m}(t) = m_0[\cos(\omega t)\,\hat{\boldsymbol{x}} + \sin(\omega t)\,\hat{\boldsymbol{y}}]. \tag{46}$$

The dipole in Fig.3 sits between a pair of non-conducting, uniformly-charged, infinitely-large parallel plates, which produce a constant and uniform electric field $\boldsymbol{E}_0 = (\sigma_0/\varepsilon_0)\hat{\boldsymbol{x}}$ in the region of space between the two plates. The distance between the plates is $d$, and the surface electrical charge-densities of the two plates are assumed to be $\pm\sigma_0$. The slow rotation of the dipole ensures that the EM radiation is negligible, and that quasi-static treatment of the EM fields is permitted. (Of course, for a magnetic dipole to rotate at a constant angular velocity, a torque must be applied to the dipole. We may assume that a constant, uniform magnetic field $H_0\hat{\boldsymbol{z}}$ is present at and around the origin of coordinates in the system of Fig.3. The presence of this magnetic field will in no way affect the ensuing analysis.)

In the Einstein-Laub formulation [33], the $E$-field exerts no torque on the dipole. However, the dipole experiences a net force $\boldsymbol{F}(t)$ in the presence of the external $E$-field. Denoting the magnetization distribution within the dipole by $\boldsymbol{M}(\boldsymbol{r},t)$, we will have

$$\boldsymbol{F}(t) = -\iiint_{-\infty}^{\infty} \partial_t \boldsymbol{M}(\boldsymbol{r},t) \times \varepsilon_0 \boldsymbol{E}_0 \, dxdydz = \sigma_0 m_0 \omega \cos(\omega t)\, \hat{\boldsymbol{z}}. \tag{47}$$

Next, we compute the EM (Abraham) momentum $\boldsymbol{p}_{em}(t)$ of the field [31-33], and show that the force $\boldsymbol{F}(t)$ is *not* rooted in an exchange of momentum between the EM field and the dipole.

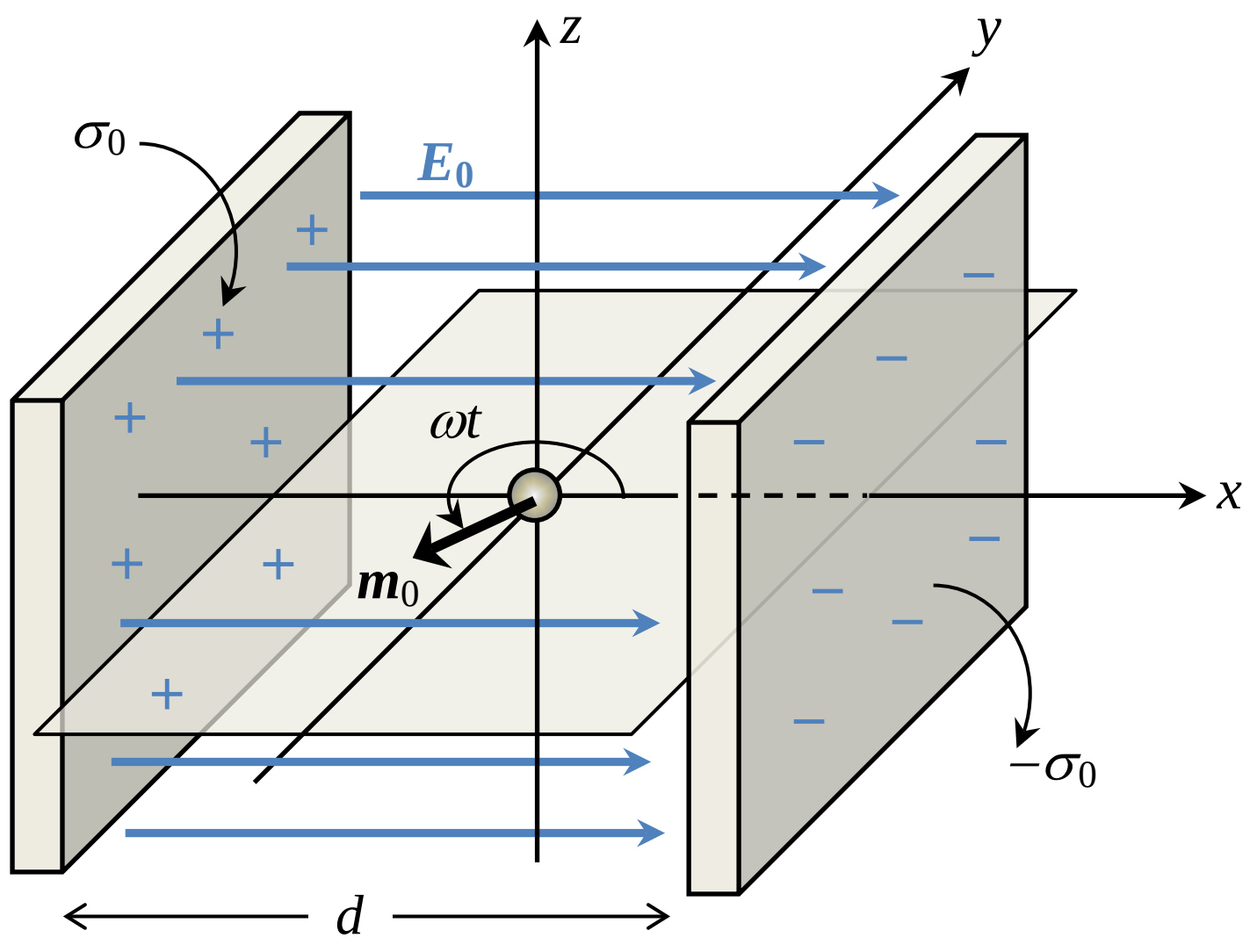

**Fig. 3**. A small magnetic dipole $\boldsymbol{m}_0$ rotates slowly in a constant, uniform $E$-field produced by a pair of uniformly-charged parallel plates. The surface electric charge densities of the plates are denoted by $\pm\sigma_0$.

Since the magnetic particle is assumed to rotate slowly, we may take the instantaneous magnetization profile $\boldsymbol{M}(\boldsymbol{r},t)$ and compute the corresponding magnetic charge-density distribution $\rho_m(\boldsymbol{r},t) = -\boldsymbol{\nabla}\cdot\boldsymbol{M}(\boldsymbol{r},t)$. The $H$-field distribution will then be obtained using Maxwell's magnetostatic equations [4-8], as follows:

$$\boldsymbol{H}(\boldsymbol{r},t) = \iiint_{-\infty}^{\infty} \frac{\rho_m(\boldsymbol{r}',t)(\boldsymbol{r}-\boldsymbol{r}')}{4\pi\mu_0|\boldsymbol{r}-\boldsymbol{r}'|^3} dx'dy'dz'. \tag{48}$$

In Eq.(48) $\mu_0$ is the vacuum permeability. In the region of space between parallel plates of Fig.3, the Abraham momentum of the EM field may be computed as follows:

$$\boldsymbol{p}_{em}(t) = \frac{1}{c^2}\int_{x=-d/2}^{+d/2}\int_{y=-\infty}^{\infty}\int_{z=-\infty}^{\infty} \boldsymbol{E}(\boldsymbol{r},t)\times\boldsymbol{H}(\boldsymbol{r},t)dxdydz$$

$$= \frac{\sigma_0}{4\pi}\iiint_{-\infty}^{\infty}\rho_m(\boldsymbol{r}',t)\left[\int_{x=-d/2}^{+d/2}\int_{y=-\infty}^{\infty}\int_{z=-\infty}^{\infty}\frac{\hat{\boldsymbol{x}}\times(\boldsymbol{r}-\boldsymbol{r}')}{|\boldsymbol{r}-\boldsymbol{r}'|^3}dxdydz\right]dx'dy'dz'$$

$$= \frac{\sigma_0}{4\pi}\iiint_{-\infty}^{\infty}\rho_m(\boldsymbol{r}',t)\left[\int_{x=-d/2}^{+d/2}\int_{y=-\infty}^{\infty}\int_{z=-\infty}^{\infty}\frac{(y-y')\hat{\boldsymbol{z}}-(z-z')\hat{\boldsymbol{y}}}{[(x-x')^2+(y-y')^2+(z-z')^2]^{3/2}}dxdydz\right]dx'dy'dz'$$

$$= \frac{\sigma_0}{4\pi}\iiint_{-\infty}^{\infty}\rho_m(\boldsymbol{r}',t)\int_{x=-d/2}^{+d/2}\left\{\int_{y=-\infty}^{\infty}\left[\cancelto{0}{\frac{\hat{\boldsymbol{y}}}{\sqrt{(x-x')^2+(y-y')^2+(z-z')^2}}\bigg|_{z=-\infty}^{\infty}}\right]dy\right.$$

$$\left.-\int_{z=-\infty}^{\infty}\left[\cancelto{0}{\frac{\hat{\boldsymbol{z}}}{\sqrt{(x-x')^2+(y-y')^2+(z-z')^2}}\bigg|_{y=-\infty}^{\infty}}\right]dz\right\}dxdx'dy'dz' = 0. \qquad (49)$$

In the absence of a net EM momentum in the region between the plates, the force $\boldsymbol{F}(t)$ of Eq.(47) cannot be balanced by a corresponding change in the field momentum $\boldsymbol{p}_{em}(t)$. We must, therefore, look elsewhere for confirmation of momentum conservation. Specifically, we note that the rotating dipole's magnetic field, being time-dependent, must give rise to an $E$-field which would then exert a force on the electrically-charged parallel plates. Since, in the absence of retardation effects, the vector potential $\boldsymbol{A}(\boldsymbol{r},t)$ of the slowly-rotating magnetic dipole is given by

$$\boldsymbol{A}(\boldsymbol{r},t) = \frac{m_0}{4\pi r^3}[\cos(\omega t)\,\hat{\boldsymbol{x}} + \sin(\omega t)\,\hat{\boldsymbol{y}}] \times \boldsymbol{r}$$

$$= \frac{m_0}{4\pi r^3}[\cos(\omega t)\,(y\hat{\boldsymbol{z}} - z\hat{\boldsymbol{y}}) + \sin(\omega t)\,(z\hat{\boldsymbol{x}} - x\hat{\boldsymbol{z}})], \qquad (50)$$

the corresponding $E$-field is readily evaluated as follows:

$$\boldsymbol{E}(\boldsymbol{r},t) = -\partial_t\boldsymbol{A}(\boldsymbol{r},t) = \frac{m_0\omega}{4\pi r^3}[\sin(\omega t)\,(y\hat{\boldsymbol{z}} - z\hat{\boldsymbol{y}}) - \cos(\omega t)\,(z\hat{\boldsymbol{x}} - x\hat{\boldsymbol{z}})]. \qquad (51)$$

To determine the force exerted by the above $E$-field on each of the charged plates, we fix the coordinate $x$ and integrate the field over the $yz$-plane to obtain

$$\iint_{-\infty}^{\infty}\boldsymbol{E}(\boldsymbol{r},t)dydz$$

$$= -\frac{m_0\omega}{4\pi}\iint_{-\infty}^{\infty}\left\{\frac{z}{r^3}[\cos(\omega t)\,\hat{\boldsymbol{x}} + \sin(\omega t)\hat{\boldsymbol{y}}] - \left[\frac{x}{r^3}\cos(\omega t) + \frac{y}{r^3}\sin(\omega t)\right]\hat{\boldsymbol{z}}\right\}dydz$$

$$= \frac{m_0\omega}{4\pi}[\cos(\omega t)\,\hat{\boldsymbol{x}} + \sin(\omega t)\hat{\boldsymbol{y}}]\int_{y=-\infty}^{\infty}\left(\cancelto{0}{\frac{1}{\sqrt{x^2+y^2+z^2}}\bigg|_{z=-\infty}^{\infty}}\right)dy$$

$$-\frac{m_0\omega}{4\pi}\sin(\omega t)\hat{\boldsymbol{z}}\int_{z=-\infty}^{\infty}\left(\cancelto{0}{\frac{1}{\sqrt{x^2+y^2+z^2}}\bigg|_{y=-\infty}^{\infty}}\right)dz + \frac{m_0\omega}{4\pi}x\cos(\omega t)\,\hat{\boldsymbol{z}}\iint_{-\infty}^{\infty}\frac{1}{r^3}dydz$$

$$= \frac{m_0\omega}{4\pi}x\cos(\omega t)\,\hat{\boldsymbol{z}}\int_0^{\infty}\frac{2\pi\rho d\rho}{(x^2+\rho^2)^{3/2}} = -½m_0\omega x\cos(\omega t)\,\hat{\boldsymbol{z}}\left(\frac{1}{\sqrt{x^2+\rho^2}}\right)\bigg|_{\rho=0}^{\infty}$$

$$= ½m_0\omega\cos(\omega t)(x/|x|)\,\hat{\boldsymbol{z}}. \qquad (52)$$

For the plate on the left-hand-side in Fig.3, $x$ is negative and $\sigma_0$ is positive, whereas for the plate on the right-hand-side, $x$ is positive while $\sigma_0$ is negative. The total force on both plates is, therefore, $\boldsymbol{F}(t) = -\sigma_0 m_0 \omega \cos(\omega t)\,\hat{\boldsymbol{z}}$. This is exactly equal in magnitude and opposite in sign to the force exerted on the magnetic dipole; see Eq.(47). Note that the force exerted by the rotating magnetic dipole on the parallel plates is independent of the separation $d$ between the plates. Consequently, the effect of the dipole on the plates cannot be ignored, even if the plates happen to be infinitely far away. We have thus confirmed the conservation of momentum by showing explicitly that the action on the dipole is equal and opposite to the reaction of the parallel plates.

**4.1. Hidden momentum of the rotating magnetic dipole**. Let us now consider the same problem from the perspective of the Lorentz formulation [1]. In this case, since the dipole has no electric charge, it does not experience any force from the $E$-field, nor does the bound-current of the dipole experience any force in the absence of an external $B$-field. So, it might be thought that, in the system of Fig.3, the Lorentz formulation predicts no forces acting on the dipole. Now, in the Lorentz formulation, the EM momentum is the Livens momentum [34], whose density is $\varepsilon_0 \boldsymbol{E}(\boldsymbol{r},t) \times \boldsymbol{B}(\boldsymbol{r},t)$. Since $\boldsymbol{B} = \mu_0 \boldsymbol{H} + \boldsymbol{M}$, and, in accordance with Eq.(49), the $\mu_0 \boldsymbol{H}$ part of $\boldsymbol{B}$ does not contribute to the EM momentum, the Livens momentum in the system of Fig.3 must be

$$\boldsymbol{p}_{em}^{(\text{Livens})} = \varepsilon_0 \boldsymbol{E}_0 \times \boldsymbol{m}(t) = \sigma_0 m_0 \sin(\omega t)\,\hat{\boldsymbol{z}}. \tag{53}$$

[The same result as in Eq.(53) may be obtained using an alternative method of calculation outlined in Appendix B.] It is thus tempting to associate the time-rate-of-change of the Livens momentum of Eq.(53) with the mechanical force exerted on the parallel plates, and to declare that momentum conservation is upheld.

However, in the Lorentz formulation there is known to be a certain amount of hidden momentum $\boldsymbol{p}_{\text{hidden}} = \boldsymbol{m}(t) \times \varepsilon_0 \boldsymbol{E}_0 = -\sigma_0\, m_0 \sin(\omega t)\,\hat{\boldsymbol{z}}$ associated with the magnetic dipole residing in the external $E$-field [29-32]. The time-rate-of-change of the hidden momentum being given by $\partial_t \boldsymbol{p}_{\text{hidden}} = -\sigma_0 m_0 \omega \cos(\omega t)\,\hat{\boldsymbol{z}}$, it is clear that an equal and opposite force must be experienced by the rotating dipole. The EM force experienced by the dipole in the Lorentz formulation, assuming the contribution of hidden momentum is properly taken into account, is thus seen to be in agreement with the result obtained in Eq.(47) via the Einstein-Laub formulation. Once again the total momentum is seen to be conserved, albeit with the inclusion of the hidden momentum and the Livens momentum in addition to the mechanical momenta imparted to the magnetic dipole and to the pair of parallel plates.

**5. Rotating electric dipole in a constant and uniform magnetic field**. In this section we examine the dual of the system of Fig.3, namely, a rotating electric dipole in a uniform magnetic field. With reference to Fig.4, consider a constant electric dipole $\boldsymbol{p}_0$, located at the origin of the coordinate system and rotating slowly at a fixed angular velocity $\omega$ in the $xz$-plane around the $y$-axis. The dipole moment may thus be written

$$\boldsymbol{p}(t) = p_0[\cos(\omega t)\,\hat{\boldsymbol{x}} + \sin(\omega t)\,\hat{\boldsymbol{z}}]. \tag{54}$$

In the presence of a constant, uniform magnetic field $\boldsymbol{H}(\boldsymbol{r},t) = H_0\hat{\boldsymbol{z}}$, produced by an infinitely-long cylinder of radius $R_c$ carrying a constant surface-current-density $\boldsymbol{J}_s = J_0\hat{\boldsymbol{\varphi}} = H_0\hat{\boldsymbol{\varphi}}$, the Lorentz force law yields the EM force exerted on the dipole as follows:

$$\boldsymbol{F}(t) = \partial_t \boldsymbol{p}(t) \times \mu_0 \boldsymbol{H} = \mu_0 H_0 p_0 \omega \sin(\omega t)\,\hat{\boldsymbol{y}}. \tag{55}$$

To confirm the conservation of momentum, we must first evaluate the EM counter-force exerted by the magnetic field of the rotating dipole on the current-carrying cylinder. Considering that the polarization density of the point-dipole is $\boldsymbol{P}(\boldsymbol{r},t)=\boldsymbol{p}(\boldsymbol{t})\delta(x)\delta(y)\delta(z)$, its bound current-density [4,5] is given by

$$\boldsymbol{J}_{\text{bound}}(t)=\partial_t\boldsymbol{P}(\boldsymbol{r},t)=-p_0\omega\delta(x)\delta(y)\delta(z)[\sin(\omega t)\,\hat{\boldsymbol{x}}-\cos(\omega t)\,\hat{\boldsymbol{z}}]. \tag{56}$$

Consequently, the vector potential $\boldsymbol{A}(\boldsymbol{r},t)$ of the dipole in the quasi-static limit may be written as

$$\boldsymbol{A}(\boldsymbol{r},t)=-\tfrac{\mu_0 p_0\omega}{4\pi r}[\sin(\omega t)\,\hat{\boldsymbol{x}}-\cos(\omega t)\,\hat{\boldsymbol{z}}]. \tag{57}$$

The magnetic field produced in the surrounding region by the (slowly) rotating dipole is thus given by

$$\boldsymbol{B}(\boldsymbol{r},t)=\mu_0\boldsymbol{H}(\boldsymbol{r},t)=\boldsymbol{\nabla}\times\boldsymbol{A}(\boldsymbol{r},t)=\tfrac{\mu_0 p_0\omega}{4\pi r^3}[(z\hat{\boldsymbol{y}}-y\hat{\boldsymbol{z}})\sin(\omega t)-(y\hat{\boldsymbol{x}}-x\hat{\boldsymbol{y}})\cos(\omega t)]. \tag{58}$$

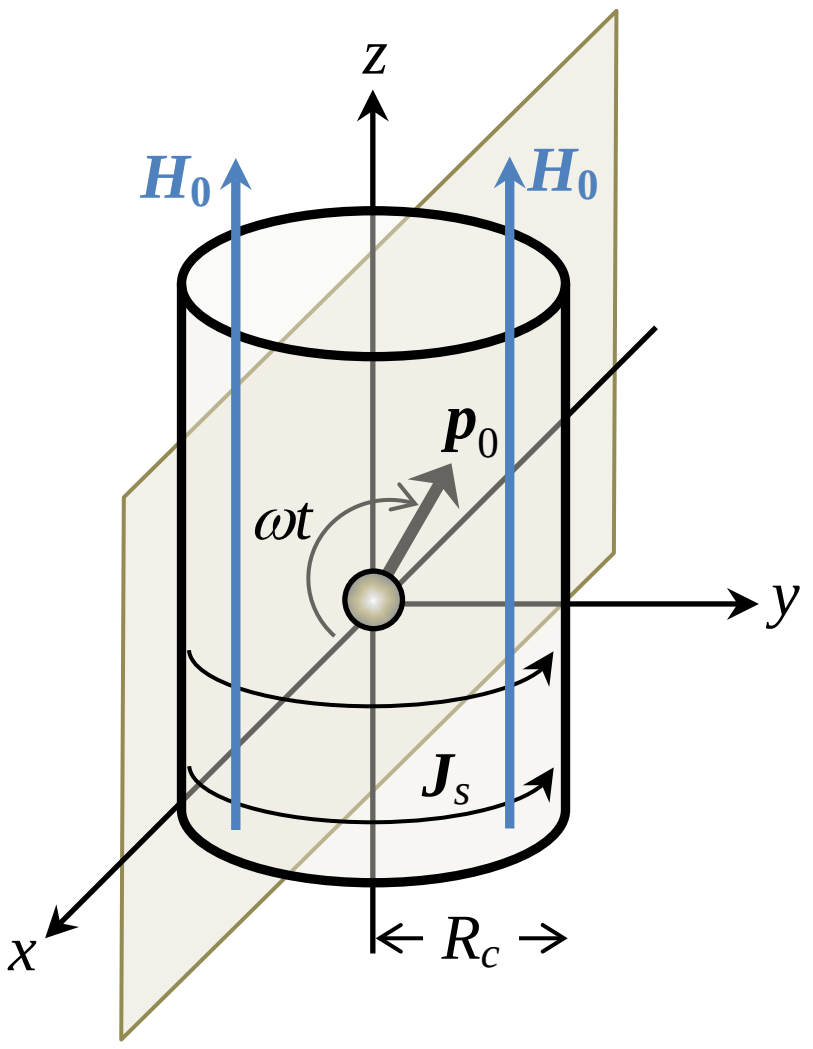

**Fig. 4**. A small spherical particle of radius $R_s$ centered at the origin of coordinates has a permanent electric dipole moment $\boldsymbol{p}_0$, which rotates slowly at a constant angular velocity $\omega$ in the $xz$-plane around the $y$-axis. The dipole is acted upon by a constant, uniform magnetic field $H_0\hat{\boldsymbol{z}}$, produced by an infinitely long, thin cylinder of radius $R_c$, which carries the constant surface-current-density $\boldsymbol{J}_s=J_0\hat{\boldsymbol{\varphi}}=H_0\hat{\boldsymbol{\varphi}}$. The dipole experiences the oscillatory Lorentz force along the $y$-axis given by Eq.(55).

We are now in a position to calculate the Lorentz force exerted by the rotating dipole on the cylinder of radius $R_c$ carrying the surface-current-density $\boldsymbol{J}_s=J_0\hat{\boldsymbol{\varphi}}=H_0\hat{\boldsymbol{\varphi}}$, which generates the magnetic field $H_0\hat{\boldsymbol{z}}$ inside the cylinder. We have

$$\begin{aligned}
\boldsymbol{F}_c(t)&=\int_{z=-\infty}^{\infty}\int_{\varphi=0}^{2\pi}J_0\hat{\boldsymbol{\varphi}}\times\mu_0\boldsymbol{H}(R_c,\varphi,z,t)R_c d\varphi dz\\
&=\int_{z=-\infty}^{\infty}\int_{\varphi=0}^{2\pi}H_0(\cos\varphi\,\hat{\boldsymbol{y}}-\sin\varphi\,\hat{\boldsymbol{x}})\times\tfrac{\mu_0 p_0\omega}{4\pi r^3}[(z\hat{\boldsymbol{y}}-y\hat{\boldsymbol{z}})\sin(\omega t)-(y\hat{\boldsymbol{x}}-x\hat{\boldsymbol{y}})\cos(\omega t)]R_c d\varphi dz\\
&=-\tfrac{\mu_0 H_0 p_0\omega R_c}{4\pi}\Big\{[\sin(\omega t)\hat{\boldsymbol{x}}-\cos(\omega t)\hat{\boldsymbol{z}}]\int_{z=-\infty}^{\infty}\int_{\varphi=0}^{2\pi}\frac{R_c\sin\varphi\cos\varphi}{(R_c^2+z^2)^{3/2}}d\varphi dz \;(\to 0)\\
&\quad+\hat{\boldsymbol{z}}\int_{z=-\infty}^{\infty}\int_{\varphi=0}^{2\pi}\frac{[x\cos(\omega t)+z\sin(\omega t)]\sin\varphi}{(R_c^2+z^2)^{3/2}}d\varphi dz\;(\to 0)+\sin(\omega t)\,\hat{\boldsymbol{y}}\int_{z=-\infty}^{\infty}\int_{\varphi=0}^{2\pi}\frac{y\sin\varphi}{(R_c^2+z^2)^{3/2}}d\varphi dz\Big\}\\
&=-\tfrac{\mu_0 H_0 p_0\omega R_c}{4\pi}\left[\int_{z=-\infty}^{\infty}\int_{\varphi=0}^{2\pi}\frac{R_c\sin^2\varphi}{(R_c^2+z^2)^{3/2}}d\varphi dz\right]\sin(\omega t)\,\hat{\boldsymbol{y}}\\
&=-¼\mu_0 H_0 p_0\omega\left[\int_{-\infty}^{\infty}\frac{d\zeta}{(1+\zeta^2)^{3/2}}\right]\sin(\omega t)\,\hat{\boldsymbol{y}}=-¼\mu_0 H_0 p_0\omega\left(\zeta/\sqrt{1+\zeta^2}\right)\Big|_{-\infty}^{\infty}\sin(\omega t)\,\hat{\boldsymbol{y}}\\
&=-½\mu_0 H_0 p_0\omega\sin(\omega t)\,\hat{\boldsymbol{y}}.
\end{aligned} \tag{59}$$

The cylinder thus experiences a force only half as large as that experienced by the dipole, and in the opposite direction; see Eq.(55). The missing momentum must, therefore, come from the EM momentum inside the volume of the cylinder. To verify this, we take the dipole to be a small spherical particle of radius $R_s$ and uniform polarization $\boldsymbol{P}(t)$. In fact, it is best to imagine two overlapping spherical dipoles, one oscillating along $x$, the other along the $z$-axis. Inside each spherical dipole, the $E$-field is uniform and given by $-\boldsymbol{P}(t)/(3\varepsilon_0)$. The EM momentum inside the spherical dipole oscillating along the $x$-axis is thus equal to $\boldsymbol{\wp}_{em}(t) = ⅓\mu_0 H_0 p_0 \cos(\omega t)\,\hat{\boldsymbol{y}}$, whereas that inside the dipole oscillating along the $z$-axis is zero. The EM momentum outside the spherical dipole oscillating along $z$ also turns out to be zero. To see this, note that the quasi-static scalar potential $\psi(\boldsymbol{r},t)$ associated with $\boldsymbol{p}(t) = p_0 \sin(\omega t)\,\hat{\boldsymbol{z}}$ in the region outside the sphere of radius $R_s$ is given by [4-8]

$$\psi(\boldsymbol{r},t) = \frac{p_0 \sin(\omega t) z}{4\pi\varepsilon_0 r^3} = \frac{p_0 \sin(\omega t) z}{4\pi\varepsilon_0(\rho^2 + z^2)^{3/2}}. \tag{60}$$

In the above equation, $r$ is the distance from the origin to the observation point (spherical coordinates), whereas $\rho$ is the radial distance from the $z$-axis (cylindrical coordinates). The corresponding $E$-field and EM momentum-density inside the cylinder are thus found to be

$$\boldsymbol{E}(\boldsymbol{r},t) = -\boldsymbol{\nabla}\psi(\boldsymbol{r},t) = -(\partial_\rho\psi)\hat{\boldsymbol{\rho}} - (\partial_z\psi)\hat{\boldsymbol{z}}. \tag{61}$$

$$\boldsymbol{E}(\boldsymbol{r},t) \times \boldsymbol{H}(\boldsymbol{r},t)/c^2 = -\left[\frac{3\mu_0 H_0 p_0 \sin(\omega t) z\rho}{4\pi(\rho^2 + z^2)^{5/2}}\right]\hat{\boldsymbol{\varphi}}. \tag{62}$$

Clearly, the integral of the momentum-density in Eq.(62) over the cylinder's volume (excluding the particle's spherical volume) vanishes. We conclude that the dipole oscillating along the $z$-axis does *not* contribute to the EM momentum of the system of Fig.4. As for the dipole oscillating along the $x$-axis, we have (in the region inside the cylinder and outside the spherical particle)

$$\psi(\boldsymbol{r},t) = \frac{p_0 \cos(\omega t) x}{4\pi\varepsilon_0 r^3} = \frac{p_0 \cos(\omega t)\rho\cos\varphi}{4\pi\varepsilon_0(\rho^2 + z^2)^{3/2}}. \tag{63}$$

$$\boldsymbol{E}(\boldsymbol{r},t) = -\boldsymbol{\nabla}\psi(\boldsymbol{r},t) = -(\partial_\rho\psi)\hat{\boldsymbol{\rho}} - \rho^{-1}(\partial_\varphi\psi)\hat{\boldsymbol{\varphi}} - (\partial_z\psi)\hat{\boldsymbol{z}}. \tag{64}$$

$$\begin{aligned}
\boldsymbol{E}(\boldsymbol{r},t) \times \boldsymbol{H}(\boldsymbol{r},t)/c^2 &= \mu_0\varepsilon_0 H_0\left[(\partial_\rho\psi)\hat{\boldsymbol{\varphi}} - \rho^{-1}(\partial_\varphi\psi)\hat{\boldsymbol{\rho}}\right] \\
&= \frac{\mu_0 H_0 p_0 \cos(\omega t)}{4\pi}\left\{\left[\frac{1}{(\rho^2+z^2)^{3/2}} - \frac{3\rho^2}{(\rho^2+z^2)^{5/2}}\right]\cos\varphi\,\hat{\boldsymbol{\varphi}} + \frac{\sin\varphi}{(\rho^2+z^2)^{3/2}}\hat{\boldsymbol{\rho}}\right\} \\
&= \frac{\mu_0 H_0 p_0 \cos(\omega t)}{4\pi}\left\{\left[\frac{1}{(\rho^2+z^2)^{3/2}} - \frac{3\rho^2}{(\rho^2+z^2)^{5/2}}\right]\cos\varphi\,(\cos\varphi\,\hat{\boldsymbol{y}} - \sin\varphi\,\hat{\boldsymbol{x}}) + \frac{\sin\varphi}{(\rho^2+z^2)^{3/2}}(\cos\varphi\,\hat{\boldsymbol{x}} + \sin\varphi\,\hat{\boldsymbol{y}})\right\} \\
&= \frac{\mu_0 H_0 p_0 \cos(\omega t)}{4\pi}\left[\frac{\hat{\boldsymbol{y}}}{(\rho^2+z^2)^{3/2}} + \frac{3\rho^2(\sin\varphi\cos\varphi\,\hat{\boldsymbol{x}} - \cos^2\varphi\,\hat{\boldsymbol{y}})}{(\rho^2+z^2)^{5/2}}\right].
\end{aligned} \tag{65}$$

It is now straightforward to integrate the EM momentum-density of Eq.(65) over the volume of the cylinder (excluding the interior of the spherical particle). The integral must be evaluated in two steps, first when $\rho$ ranges from $R_s$ to $R_c$, and then when $\rho$ goes from 0 to $R_s$. We will have

a) $$\frac{\mu_0 H_0 p_0 \cos(\omega t)}{4\pi}\int_{\rho=R_s}^{R_c}\int_{\varphi=0}^{2\pi}\int_{z=-\infty}^{\infty}\left[\frac{\hat{\boldsymbol{y}}}{(\rho^2+z^2)^{3/2}} + \frac{3\rho^2(\sin\varphi\cos\varphi\,\hat{\boldsymbol{x}} - \cos^2\varphi\,\hat{\boldsymbol{y}})}{(\rho^2+z^2)^{5/2}}\right]\rho d\rho d\varphi dz$$

$$= \mu_0 H_0 p_0 \cos(\omega t)\,\hat{\boldsymbol{y}}\int_{\rho=R_s}^{R_c}\int_{z=0}^{\infty}\left[\frac{\rho}{(\rho^2+z^2)^{3/2}} - \frac{3\rho^3}{2(\rho^2+z^2)^{5/2}}\right]d\rho dz$$ ← See Appendix A

$$= \mu_0 H_0 p_0 \cos(\omega t)\,\hat{\boldsymbol{y}} \int_{\rho=R_s}^{R_c} \left(\frac{1}{\rho} - \frac{1}{\rho}\right) d\rho = 0. \tag{66a}$$

b) $\mu_0 H_0 p_0 \cos(\omega t)\,\hat{\boldsymbol{y}} \int_{\rho=0}^{R_s} \int_{z=(R_s^2-\rho^2)^{½}}^{\infty} \left[\frac{\rho}{(\rho^2+z^2)^{3/2}} - \frac{3\rho^3}{2(\rho^2+z^2)^{5/2}}\right] d\rho dz$ ← See Appendix A

$$= \mu_0 H_0 p_0 \cos(\omega t)\,\hat{\boldsymbol{y}} \int_0^{R_s} \left\{\frac{1}{\rho}\left[1 - \frac{(R_s^2-\rho^2)^{1/2}}{R_s}\right] - \frac{3}{2\rho}\left[1 - \frac{(R_s^2-\rho^2)^{1/2}}{R_s} - \frac{1}{3} + \frac{(R_s^2-\rho^2)^{3/2}}{3R_s^3}\right]\right\} d\rho$$

$$= ½\mu_0 H_0 p_0 \cos(\omega t)\,\hat{\boldsymbol{y}} \int_0^{R_s} \left(\rho\sqrt{R_s^2 - \rho^2}/R_s^3\right) d\rho$$

$$= ½\mu_0 H_0 p_0 \cos(\omega t)\,\hat{\boldsymbol{y}} \int_0^1 \eta\sqrt{1-\eta^2}\, d\eta = \tfrac{1}{6}\mu_0 H_0 p_0 \cos(\omega t)\,\hat{\boldsymbol{y}}. \tag{66b}$$ ← See Appendix A

The total EM momentum (including the contribution from the interior region of the spherical dipole) is thus seen to be $\boldsymbol{\wp}_{em}(t) = ½\mu_0 H_0 p_0 \cos(\omega t)\,\hat{\boldsymbol{y}}$, confirming the conservation of overall momentum (i.e., electromagnetic plus mechanical) in the system of Fig.4.

**6. Energy of a dipole in an external EM field**. Our next example concerns the EM energy of an electric dipole in a constant and uniform external electric field. The $E$-field of a small, uniformly-polarized spherical particle (i.e., spherical dipole) located at the origin of a spherical coordinate system and oriented along the $z$-axis is given by [4-8]

$$\boldsymbol{E}(\boldsymbol{r}) = \begin{cases} \dfrac{p_0(2\cos\theta\,\hat{\boldsymbol{r}} + \sin\theta\,\hat{\boldsymbol{\theta}})}{4\pi\varepsilon_0 r^3}, & r \ge R \\ -\dfrac{p_0\hat{\boldsymbol{z}}}{4\pi\varepsilon_0 R^3}, & r < R. \end{cases} \tag{67}$$

Here $R$ is the radius of the spherical particle and $p_0\hat{\boldsymbol{z}}$ is its (electric) dipole moment. In a static external electric field $\boldsymbol{E}^{(\text{ext})}(\boldsymbol{r})$, the energy of the dipole is said to be

$$\mathcal{E} = -\boldsymbol{p}\cdot\boldsymbol{E}^{(\text{ext})}(0) = -p_0 E_z^{(\text{ext})}(0). \tag{68}$$

This is because the torque exerted by $\boldsymbol{E}^{(\text{ext})}(\boldsymbol{r})$ on the dipole $\boldsymbol{p}$ oriented in an arbitrary direction is given by $\boldsymbol{\tau} = \boldsymbol{p}\times\boldsymbol{E}^{(\text{ext})}(0)$. Denoting the angle between $\boldsymbol{p}$ and $\boldsymbol{E}^{(\text{ext})}(0)$ by $\varphi$, the magnitude of the torque is readily seen to be $p_0 E^{(\text{ext})}(0)\sin\varphi$. If now the angle $\varphi$ changes by $\delta\varphi$, the work done by the $E$-field on the dipole will be

$$\delta W = p_0 E^{(\text{ext})}(0)\sin\varphi\,\delta\varphi = -\delta\left[p_0 E^{(\text{ext})}(0)\cos\varphi\right] = -\delta\left[\boldsymbol{p}\cdot\boldsymbol{E}^{(\text{ext})}(0)\right]. \tag{69}$$

The energy of the dipole in the external $E$-field is thus said to be $-\boldsymbol{p}\cdot\boldsymbol{E}^{(\text{ext})}(0)$, with the zero of energy assigned (arbitrarily) to that orientation of the dipole which is perpendicular to $\boldsymbol{E}^{(\text{ext})}(0)$. This energy is exchanged between the dipole's rotational kinetic energy and the stored energy in the $E$-field in the region of space surrounding the dipole. To see this, note that the energy-density of the $E$-field is generally given by [4,5]

$$\mathcal{E}(\boldsymbol{r}) = ½\varepsilon_0|\boldsymbol{E}_{\text{total}}(\boldsymbol{r})|^2 = ½\varepsilon_0\left|\boldsymbol{E}(\boldsymbol{r}) + \boldsymbol{E}^{(\text{ext})}(\boldsymbol{r})\right|^2$$

$$= ½\varepsilon_0|\boldsymbol{E}(\boldsymbol{r})|^2 + ½\varepsilon_0\left|\boldsymbol{E}^{(\text{ext})}(\boldsymbol{r})\right|^2 + \varepsilon_0\boldsymbol{E}(\boldsymbol{r})\cdot\boldsymbol{E}^{(\text{ext})}(\boldsymbol{r}). \tag{70}$$

The first two terms on the right-hand-side of the above equation are independent of the orientation of the dipole. Thus the stored energy in the $E$-field varies as the integral of the third term, $\varepsilon_0 \boldsymbol{E}(\boldsymbol{r}) \cdot \boldsymbol{E}^{(\text{ext})}(\boldsymbol{r})$, over the entire space.

When the dipole is aligned with the $z$-axis, its $E$-field given by Eq.(67) maybe equivalently expressed as follows:

$$\boldsymbol{E}(\boldsymbol{r}) = -\frac{p_0}{4\pi\varepsilon_0}\frac{\partial}{\partial z}\left(\frac{\boldsymbol{r}}{r^3}\right). \tag{71}$$

This should be obvious, considering that the dipole consists of a pair of point-charges, $\pm q$, separated by a short distance $d$ along the $z$-axis. The $E$-field produced by a charge $q$ located at the origin of the coordinate system is given by $q\boldsymbol{r}/(4\pi\varepsilon_0 r^3)$. Separating the $\pm q$ charges by a short distance $d$ along the $z$-axis and recalling that $p_0 = qd$ yields the dipolar $E$-field of Eq.(71). Note, however, that Eq.(71) does *not* exactly reproduce the internal field of the dipole in the region $r < R$ as specified in Eq.(67). Nevertheless, in the vicinity of the origin, the general behavior of the $E$-field of Eq.(71) is consistent with that of the spherical dipole described by Eq.(67).

The relevant part of the $E$-field energy in Eq.(70), the third term on the right-hand-side (i.e., the cross-term), may now be written as follows:

$$\begin{aligned}\Delta\mathcal{E} &= \iiint_{-\infty}^{\infty} \varepsilon_0 \boldsymbol{E}(\boldsymbol{r}) \cdot \boldsymbol{E}^{(\text{ext})}(\boldsymbol{r}) dxdydz = -\frac{p_0}{4\pi}\iiint_{-\infty}^{\infty} \frac{\partial}{\partial z}\left(\frac{\boldsymbol{r}}{r^3}\right) \cdot \boldsymbol{E}^{(\text{ext})}(\boldsymbol{r}) dxdydz \\ &= \frac{p_0}{4\pi}\iiint_{-\infty}^{\infty} \frac{\boldsymbol{r}}{r^3} \cdot \frac{\partial \boldsymbol{E}^{(\text{ext})}(\boldsymbol{r})}{\partial z} dxdydz. \end{aligned} \tag{72}$$

The final step in the above derivation involves an integration by parts followed by the assumption that the external $E$-field at $\pm\infty$ is sufficiently weak to be negligible. We now invoke the static nature of $\boldsymbol{E}^{(\text{ext})}(\boldsymbol{r})$ and the fact that $\boldsymbol{\nabla} \times \boldsymbol{E}^{(\text{ext})}(\boldsymbol{r}) = 0$ to replace $\partial_z E_y^{(\text{ext})}$ by $\partial_y E_z^{(\text{ext})}$ and $\partial_z E_x^{(\text{ext})}$ by $\partial_x E_z^{(\text{ext})}$. We will have

$$\begin{aligned}\Delta\mathcal{E} &= \frac{p_0}{4\pi}\iiint_{-\infty}^{\infty} \frac{\boldsymbol{r}}{r^3} \cdot \left[\partial_x E_z^{(\text{ext})} \hat{\boldsymbol{x}} + \partial_y E_z^{(\text{ext})} \hat{\boldsymbol{y}} + \partial_z E_z^{(\text{ext})} \hat{\boldsymbol{z}}\right] dxdydz \\ &= -\frac{p_0}{4\pi}\iiint_{-\infty}^{\infty} \left[\boldsymbol{\nabla} \cdot \left(\frac{\boldsymbol{r}}{r^3}\right)\right] E_z^{(\text{ext})}(\boldsymbol{r}) dxdydz \\ &= -\frac{p_0}{4\pi}\iiint_{-\infty}^{\infty} 4\pi\delta(\boldsymbol{r}) E_z^{(\text{ext})}(\boldsymbol{r}) dxdydz = -p_0 E_z^{(\text{ext})}(0). \end{aligned} \tag{73}$$

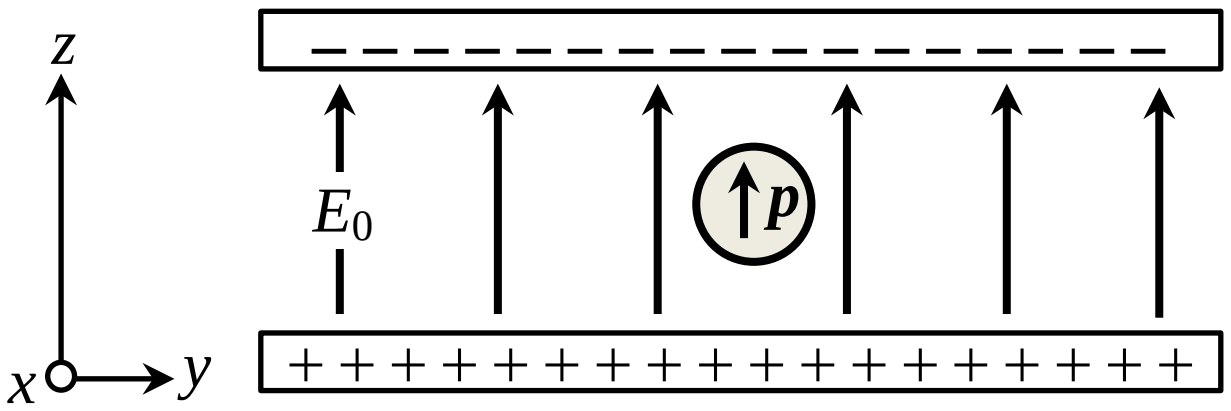

**Fig.5**. A small spherical dipole $\boldsymbol{p} = p_0 \hat{\boldsymbol{z}}$ is placed in the uniform $E$-field between two uniformly-charged parallel plates. The EM energy stored in the $E$-field contains a contribution from the cross-term $\varepsilon_0 \boldsymbol{E}(\boldsymbol{r}) \cdot \boldsymbol{E}^{(\text{ext})}(\boldsymbol{r})$, where $\boldsymbol{E}^{(\text{ext})}(\boldsymbol{r}) = E_0 \hat{\boldsymbol{z}}$.

Thus, when the dipole is anti-parallel to $\boldsymbol{E}^{(\text{ext})}(0)$, the energy stored in the $E$-field is at its peak value. Allowing the dipole to rotate will enable the torque to bring it into alignment with the local external field. In the process the dipole acquires rotational kinetic energy while the energy stored in the $E$-field is reduced by an equal amount. A simple scenario involving a dipole $\boldsymbol{p}$ placed between two uniformly-charged, infinitely large, parallel plates, as shown in Fig.5, would be instructive. It is easy to verify by direct integration that, when $\boldsymbol{p} = p_0\hat{\boldsymbol{z}}$ and the dipolar $E$-field $\boldsymbol{E}(\boldsymbol{r})$ is given by Eq.(67) while $\boldsymbol{E}^{(\text{ext})}(\boldsymbol{r}) = E_0\hat{\boldsymbol{z}}$ in the region between the parallel plates, the cross-term $\varepsilon_0\boldsymbol{E}(\boldsymbol{r}) \cdot \boldsymbol{E}^{(\text{ext})}(\boldsymbol{r})$ appearing in Eq.(70) would integrate to $-p_0E_0$.

**6.1. The case of a relativistically-induced electric dipole**. Next, with reference to Fig. 6, consider a small, uniformly-magnetized, spherical particle (i.e., a magnetic dipole) located at the origin of the $x'y'z'$ coordinate system, with its magnetization aligned with the $z'$-axis. The magnetic field of the dipole is given by

$$\boldsymbol{H}(\boldsymbol{r}') = \begin{cases} \dfrac{m_0(2\cos\theta'\hat{\boldsymbol{r}} + \sin\theta'\hat{\boldsymbol{\theta}})}{4\pi\mu_0 r'^3}, & r' \geq R \\ -\dfrac{m_0\hat{\boldsymbol{z}}}{4\pi\mu_0 R^3}, & r' < R. \end{cases} \tag{74}$$

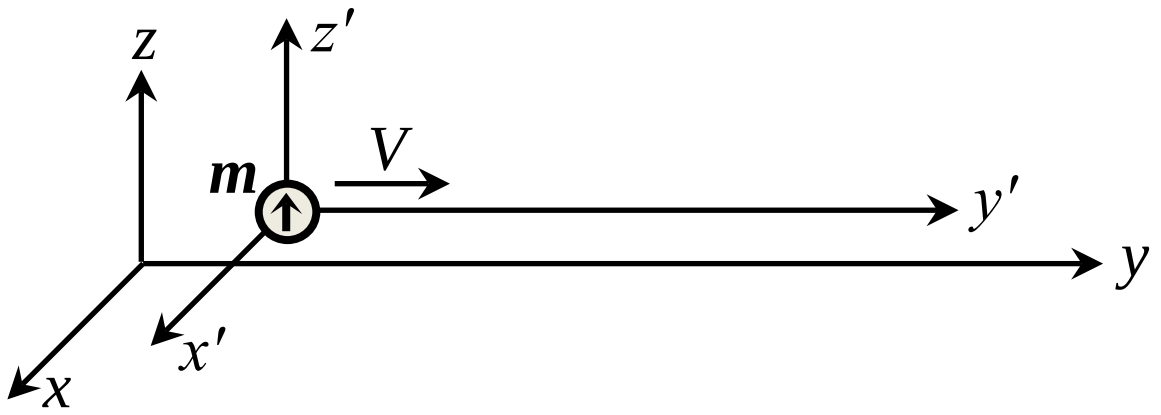

**Fig. 6**. A small spherical magnetic dipole $\boldsymbol{m} = m_0\hat{\boldsymbol{z}}$ moves with constant velocity $V$ along the $y$-axis of a Cartesian coordinate system. The rest-frame of the particle is identified as $x'y'z'$.

Note that, in order to maintain complete symmetry with the case of an electric dipole, we are defining the magnetic dipole moment $m_0\hat{\boldsymbol{z}}$ and its corresponding magnetization $\boldsymbol{M} = 3m_0\hat{\boldsymbol{z}}/(4\pi R^3)$ such that, in the $SI$ system of units, $\boldsymbol{B} = \mu_0\boldsymbol{H} + \boldsymbol{M}$. In this way, the magnetic dipole moment of a small loop of area $A$ carrying a constant electric current $I$ in the $xy$-plane will be $\boldsymbol{m}_0 = \mu_0 IA\,\hat{\boldsymbol{z}}$.

Suppose now that the $x'y'z'$ frame moves with constant velocity $V$ along the $y$-axis. From the perspective of an observer in the stationary frame $xyz$ (i.e., the lab frame), the magnetic dipole $m_0\hat{\boldsymbol{z}}$ will be accompanied by a relativistically-induced electric dipole $\boldsymbol{p} = \varepsilon_0 V m_0\hat{\boldsymbol{x}}$. The $E$-field in the $xyz$ frame is readily obtained from a Lorentz transformation of the $B$-field in the $x'y'z'$ frame, as follows:

$$\boldsymbol{E}(\boldsymbol{r}, t=0) = \begin{cases} -\gamma V B_z\hat{\boldsymbol{x}} + \gamma V B_x\hat{\boldsymbol{z}} = \dfrac{\gamma V m_0}{4\pi(x^2+\gamma^2y^2+z^2)^{5/2}}[(x^2 + \gamma^2y^2 - 2z^2)\hat{\boldsymbol{x}} + 3xz\hat{\boldsymbol{z}}], & r' \geq R \\ -\gamma V B_z\hat{\boldsymbol{x}} = -\dfrac{\gamma V m_0}{2\pi R^3}\hat{\boldsymbol{x}}, & r' < R. \end{cases} \tag{75}$$

Clearly, this field is very different from the $E$-field produced by an ordinary electric dipole $\boldsymbol{p} = \varepsilon_0 V m_0\hat{\boldsymbol{x}}$ placed at the origin of the $xyz$ coordinate system. In the presence of a static external $E$-field $\boldsymbol{E}^{(\text{ext})}(\boldsymbol{r})$, the $E$-field energy associated with the third term on the right-hand-side of Eq.(70) will be

$$\begin{aligned} \Delta\mathcal{E}(t=0) &= \iiint_{-\infty}^{\infty} \varepsilon_0\boldsymbol{E}(x,y,z,t=0) \cdot \boldsymbol{E}^{(\text{ext})}(x,y,z)dxdydz \\ &= \varepsilon_0\gamma^{-1}\iiint_{-\infty}^{\infty} \boldsymbol{E}(x,\gamma^{-1}y,z,t=0) \cdot \boldsymbol{E}^{(\text{ext})}(x,\gamma^{-1}y,z)dxdydz. \end{aligned} \tag{76}$$

It is readily seen from Eq.(75) that, in the region $r' \geq R$, the dipolar $E$-field may be expressed as

$$\boldsymbol{E}(x,\gamma^{-1}y,z,t=0) = \frac{\gamma V m_0}{4\pi}\frac{\partial}{\partial z}\left(\frac{z\hat{\boldsymbol{x}} - x\hat{\boldsymbol{z}}}{r^3}\right). \tag{77}$$

The above expression, however, does *not* reproduce the correct behavior for the internal $E$-field of the dipole (i.e., in the region $r' < R$), as given by Eq.(75). The reason being that $\partial_z(\boldsymbol{r}/r^3)$ is an appropriate representation for the $H$-field of the magnetic dipole, whereas the $E$-field of the induced electric dipole given by Eq.(75) is intimately tied to the $B$-field of the magnetic dipole. We thus proceed to evaluate $\Delta\mathcal{E}(t=0)$ of Eq.(76) by ignoring the difference between the $E$-field of Eq.(77) and the true internal field of the dipole. The result will subsequently be adjusted to accommodate the correction to the internal field. Substituting Eq.(77) into Eq.(76), we find

$$\begin{aligned}\Delta\mathcal{E}(t=0) &= \frac{\varepsilon_0 V m_0}{4\pi}\iiint_{-\infty}^{\infty}\frac{\partial}{\partial z}\left(\frac{z\hat{\boldsymbol{x}}-x\hat{\boldsymbol{z}}}{r^3}\right)\cdot \boldsymbol{E}^{(\text{ext})}(x,\gamma^{-1}y,z)dxdydz\\ &= -\frac{\varepsilon_0 V m_0}{4\pi}\iiint_{-\infty}^{\infty}\left(\frac{z\hat{\boldsymbol{x}}-x\hat{\boldsymbol{z}}}{r^3}\right)\cdot\left(\frac{\partial E_x^{(\text{ext})}}{\partial z}\hat{\boldsymbol{x}}+\frac{\partial E_z^{(\text{ext})}}{\partial z}\hat{\boldsymbol{z}}\right)dxdydz\\ &= -\frac{\varepsilon_0 V m_0}{4\pi}\iiint_{-\infty}^{\infty}\left(\frac{z\hat{\boldsymbol{x}}-x\hat{\boldsymbol{z}}}{r^3}\right)\cdot\left(\frac{\partial E_z^{(\text{ext})}}{\partial x}\hat{\boldsymbol{x}}+\frac{\partial E_z^{(\text{ext})}}{\partial z}\hat{\boldsymbol{z}}\right)dxdydz\\ &= \frac{\varepsilon_0 V m_0}{4\pi}\iiint_{-\infty}^{\infty}\left[\frac{\partial}{\partial x}\left(\frac{z}{r^3}\right)-\frac{\partial}{\partial z}\left(\frac{x}{r^3}\right)\right]E_z^{(\text{ext})}(x,\gamma^{-1}y,z)dxdydz = 0. \qquad (78)\end{aligned}$$

We must now add to the above result the correction due to the internal $E$-field of the induced electric dipole $\boldsymbol{p} = \varepsilon_0 V m_0\hat{\boldsymbol{x}}$. This correction arises from the difference between the internal $B$-field of the magnetic dipole and its corresponding $\mu_0 H$-field, which is just the magnetization $M$ of the dipole. In the spherical dipole model, the correction amounts to an additional $\boldsymbol{E} = -\gamma V M\hat{\boldsymbol{x}}$ in the region $r' < R$, whose contribution to $\Delta\mathcal{E}(t=0)$ is readily seen to be $-\boldsymbol{p}\cdot\boldsymbol{E}^{(\text{ext})}(0)$. In the end, we recover the standard formula for the energy of an electric dipole immersed in a static external $E$-field, even though the field distribution of a relativistically-induced electric dipole turned out to be quite different from that of an ordinary electric dipole.

**6.2. Magnetic dipole in an external magnetic field**. It is instructive to examine the similarities and differences between an electric dipole immersed in an external electric field, and a magnetic dipole immersed in an external magnetic field. One major difference is that, while in the preceding derivations involving a static $E$-field, we could use the identity $\boldsymbol{\nabla}\times\boldsymbol{E}^{(\text{ext})}(\boldsymbol{r}) = 0$, the corresponding identity cannot be invoked for magnetic fields, where $\boldsymbol{\nabla}\times\boldsymbol{H}^{(\text{ext})}(\boldsymbol{r}) = \boldsymbol{J}_{\text{free}}(\boldsymbol{r})$. As a simple example, consider a small, uniformly-magnetized spherical particle (i.e., a magnetic dipole) placed inside a pair of infinitely-long, uniformly-charged concentric cylinders whose radii are nearly equal and have equal but opposite surface-charge-densities, as shown in Fig. 7.

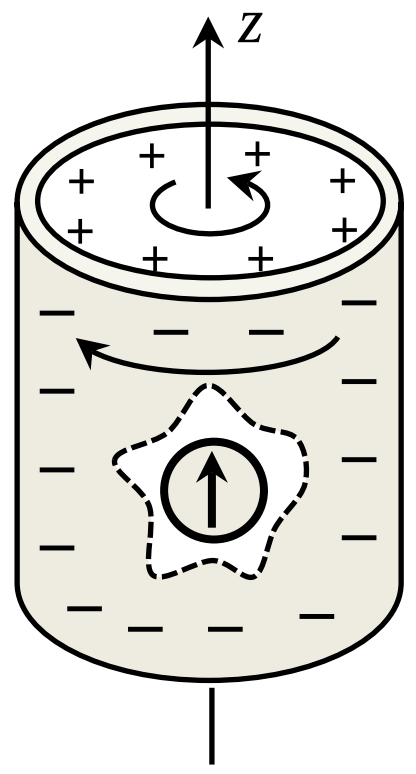

**Fig. 7**. A small spherical magnetic dipole $\boldsymbol{m} = m_0\hat{\boldsymbol{z}}$ inside a pair of concentric, infinitely-long, uniformly-charged cylinders. The electric charge-densities of the two cylinders are equal in magnitude and opposite in sign. When the cylinders are spun in opposite directions, a uniform magnetic field $\boldsymbol{H}(\boldsymbol{r}) = H_0\hat{\boldsymbol{z}}$ is created inside the cylinders.

The cylinders are initially at rest, but, at $t = 0$, they begin to rotate slowly in opposite directions. A uniform magnetic field $H_0\hat{\boldsymbol{z}}$ thus builds up inside the cylinders, immersing the magnetic dipole $m_0\hat{\boldsymbol{z}}$ in an external magnetic field. Let us first analyze the system in accordance with the Einstein-Laub formulation. Here no energy is exchanged between the dipole and the cylinders as the latter spin up. Moreover, the *H*-field inside the cylinders, being the sum of the dipole field and the uniform field set up by the rotating cylinders, does *not* make a net contribution to the energy (via a cross-term) upon integrating the *H*-field energy-density over the interior volume of the cylinders. In other words, given the dipolar field

$$\boldsymbol{H}(\boldsymbol{r}) = \begin{cases} \dfrac{m_0(2\cos\theta\hat{\boldsymbol{r}} + \sin\theta\hat{\boldsymbol{\theta}})}{4\pi\mu_0 r^3}, & r \geq R \\ -\dfrac{m_0\hat{\boldsymbol{z}}}{4\pi\mu_0 R^3}, & r < R, \end{cases} \tag{79}$$

the total energy-density inside the cylinders will be

$$\begin{aligned} \mathcal{E}(\boldsymbol{r}) &= ½\mu_0|\boldsymbol{H}_{\text{total}}(\boldsymbol{r})|^2 = ½\mu_0\left|\boldsymbol{H}^{(\text{ext})}(\boldsymbol{r}) + \boldsymbol{H}(\boldsymbol{r})\right|^2 \\ &= ½\mu_0 H_0^2 + ½\mu_0|\boldsymbol{H}(\boldsymbol{r})|^2 + \mu_0 H_0 H_z(\boldsymbol{r}). \end{aligned} \tag{80}$$

Now, the third term on the right-hand-side of Eq.(80) integrates to zero over the volume of the cylinder. An easy way to see this is to note that, for a point-dipole $m_0\hat{\boldsymbol{z}}$, the $H$-field is

$$\boldsymbol{H}(\boldsymbol{r}) = -\frac{m_0}{4\pi\mu_0}\frac{\partial}{\partial z}\left(\frac{\boldsymbol{r}}{r^3}\right), \tag{81}$$

whose integral over $z$ from $-\infty$ to $+\infty$ ends up being zero. The question remains, however, as to what happens to the dipole's energy in the external field $H_0\hat{\boldsymbol{z}}$, because surely there will be a torque acting on the dipole if it is turned away from the *z*-axis, and the action of this torque on the rotating dipole should change its rotational kinetic energy in exactly the same way that an external $E$-field exchanges energy with a rotating electric dipole.

To answer the above question, imagine the magnetic dipole losing its magnetization, say, as a result of warming up slowly to beyond its Curie temperature. The changing magnetization produces an $E$-field in the surrounding space in accordance with $\boldsymbol{\nabla}\times\boldsymbol{E} = -\partial_t\boldsymbol{B}$, which acts on the rotating cylinders to speed them up, if the initial magnetization is pointing up — or to slow them down, if the initial magnetization is downward. The kinetic energy thus imparted to the cylinders can be shown to be exactly equal to $\pm m_0 H_0$. The energy of a magnetic dipole in an external *H*-field is thus seen to be $\mathcal{E} = -\boldsymbol{m}_0\cdot\boldsymbol{H}^{(\text{ext})}(0)$, as expected from an argument based on the torque exerted by the external *H*-field on the dipole. In contrast to the case of an external *E*-field acting on an electric dipole, however, the energy associated with a magnetic dipole is *not* exchanged with the surrounding *H*-field but, rather, is exchanged with the source of the external *H*-field, which, in the present example, is the rotating pair of cylinders.

In the Lorentz formulation, the magnetic field energy-density is $\mathcal{E}(\boldsymbol{r},t) = ½\mu_0^{-1}|\boldsymbol{B}(\boldsymbol{r},t)|^2$. In this case the stored energy-density in the cross-term, namely, $\mu_0^{-1}\boldsymbol{B}^{(\text{ext})}(\boldsymbol{r})\cdot\boldsymbol{B}(\boldsymbol{r})$, integrates to $m_0 H_0$. However, the bound current-density associated with the magnetic dipole, i.e., $\mu_0^{-1}\boldsymbol{\nabla}\times\boldsymbol{M}$, now exchanges energy with the $E$-field produced by the cylinders as they spin up from rest to their final rotational speed. (Note: In the Einstein-Laub formulation, the bound current of the magnetic dipole does *not* participate in the exchange of energy with the fields.) The energy thus extracted from the bound current during the spin-up process turns out to be exactly equal to $m_0 H_0$. Therefore, no net energy is given by the spinning cylinders to the dipole and its *B*-field during spin-up. Once again, any energy exchanged in consequence of a rotation of the magnetic dipole will be between the dipole and the spinning cylinders.

**7. Concluding remarks**. While the conservation laws of EM energy and linear and angular momenta, being direct consequences of the energy-momentum stress tensor formulation(s) of classical electrodynamics [1], are universal, their verification in specific situations may or may not be straightforward. In this paper, we have examined a few non-trivial EM systems which are sufficiently simple to be amenable to exact solutions of Maxwell's equations in conjunction with the electrodynamic laws of force, torque, energy, and momentum.

It is well known that a light pulse reflecting from a mirror or passing through an absorptive or transparent object exchanges its EM energy and momentum with the material medium. The source of the radiation does not participate in these exchanges, since the light pulse, once detached from its source, is essentially autonomous. The same, however, cannot be said about exchanges that take place between static (or quasi-static) fields and material media. As the examples of the preceding sections have amply demonstrated, the (undetached) sources of the EM fields in such situations could actively participate in the exchange of energy and momentum.

When magnetic materials interact with an external electric field, the standard (i.e., Lorentz) formulation of classical electrodynamics could deviate from an alternative formulation of force, torque, energy and momentum first proposed by A. Einstein and J. Laub in 1908 [33]. In such cases we have analyzed the system under consideration from both perspectives, so that the reader might appreciate the similarities and differences of the two formulations. There exist other formulations of classical electrodynamics (e.g., Chu, Minkowski, Abraham), each built upon the foundation of Maxwell's macroscopic equations, but deviating from the standard formulation in their structure of the corresponding energy-momentum stress tensor [34]. We have chosen to stay away from these alternative formulations, as we believe them to be phenomenological rather than rooted in realistic microscopic models of interaction between matter and EM fields.

## Appendix A

Below we list (or evaluate) the various integrals and identities used throughout the paper.

$$J_{m-1}(x)+J_{m+1}(x)=2(m/x)J_m(x) \qquad \text{(Gradshteyn \& Ryzhik* 8.471-1)} \qquad \text{(A1)}$$

$$J_{m-1}(x)-J_{m+1}(x)=2J'_m(x) \qquad \text{(Gradshteyn \& Ryzhik 8.471-2)} \qquad \text{(A2)}$$

$$(m/x)\,J_m(x)-J'_m(x)=J_{m+1}(x) \qquad \text{(Gradshteyn \& Ryzhik 8.472-2)} \qquad \text{(A3)}$$

$$\int xJ_m^2(x)dx=\tfrac{1}{2}x^2[J_m^2(x)-J_{m-1}(x)J_{m+1}(x)] \qquad \text{(Gradshteyn \& Ryzhik 5.54-2)} \qquad \text{(A4)}$$

$$\int_0^{x_0} x^2J_m(x)J'_m(x)dx=\tfrac{1}{2}x_0^2\,J_m^2(x_0)-\int_0^{x_0} xJ_m^2(x)dx=\tfrac{1}{2}x_0^2\,J_{m-1}(x_0)J_{m+1}(x_0). \qquad \text{(A5)}$$

$$\begin{aligned}\int_0^{x_0} x\big[(m/x)^2\,J_m^2(x)+{J'_m}^2(x)\big]dx\\
&=\int_0^{x_0}\{x[(m/x)\,J_m(x)-J'_m(x)]^2+2mJ'_m(x)\,J_m(x)\}dx\\
&=\int_0^{x_0}[x\,J_{m+1}^2(x)+2mJ'_m(x)\,J_m(x)]dx\\
&=\tfrac{1}{2}x_0^2\{J_{m+1}^2(x_0)-J_m(x_0)J_{m+2}(x_0)+2m[J_m(x_0)/x_0]^2\}.\end{aligned} \qquad \text{(A6)}$$

*I. S. Gradshteyn and I. M. Ryzhik, "*Table of Integrals, Series, and Products*," 7th Edition, Academic Press (2007).

$$\int_0^{x_0}[xJ_m^2(x) - xJ_{m+1}^2(x) - 2m\,J_m(x)\,J_m'(x)]dx$$
$$= ½x_0^2\{J_m^2(x_0) - J_{m-1}(x_0)J_{m+1}(x_0) - J_{m+1}^2(x_0) + J_m(x_0)J_{m+2}(x_0) - 2m[J_m(x_0)/x_0]^2\}$$
$$= ½x_0^2\{J_m(x_0)[J_m(x_0) + J_{m+2}(x_0)] - J_{m+1}(x_0)[J_{m-1}(x_0) + J_{m+1}(x_0)] - 2m[J_m(x_0)/x_0]^2\}$$
$$= x_0^2\{(m+1)J_m(x_0)J_{m+1}(x_0)/x_0 - mJ_{m+1}(x_0)J_m(x_0)/x_0 - m[J_m(x_0)/x_0]^2\}$$
$$= x_0^2\{J_m(x_0)J_{m+1}(x_0)/x_0 - m[J_m(x_0)/x_0]^2\}$$
$$= [x_0\,J_{m+1}(x_0) - mJ_m(x_0)]\,J_m(x_0). \qquad \text{(A7)}$$

$$\int \frac{dx}{\sqrt{a+x^2}} = \ln\left(\sqrt{a+x^2} + x\right) \qquad \text{(Gradshteyn \& Ryzhik 2.261)} \qquad \text{(A8)}$$

$$\int \frac{dx}{(a+x^2)^{3/2}} = \frac{x}{a\sqrt{a+x^2}} \qquad \text{(Gradshteyn \& Ryzhik 2.271-5)} \qquad \text{(A9)}$$

$$\int \frac{dx}{(a+x^2)^{5/2}} = \frac{1}{a^2}\left[\frac{x}{\sqrt{a+x^2}} - \frac{x^3}{3(a+x^2)^{3/2}}\right] \qquad \text{(Gradshteyn \& Ryzhik 2.271-6)} \qquad \text{(A10)}$$

$$\int_0^1 x\sqrt{1-x^2}dx = -⅓(1-x^2)^{3/2}\Big|_0^1 = ⅓ \qquad \text{(Gradshteyn \& Ryzhik 2.262-2)} \qquad \text{(A11)}$$

$$\int_0^{\pi} \cos x\ \ln(1 - 2a\cos x + a^2)\,dx = \begin{cases} -\pi a, & (a^2 < 1) \\ -\pi/a, & (a^2 > 1) \end{cases} \qquad \text{(Gradshteyn \& Ryzhik 4.397-6)} \qquad \text{(A12)}$$

# Appendix B

An alternative method of computing the Livens momentum for the system of Fig.3 begins by expressing the bound current-density associated with the magnetization distribution as $\boldsymbol{J}_{\text{bound}}(\boldsymbol{r},t) = \mu_0^{-1}\boldsymbol{\nabla}\times\boldsymbol{M}(\boldsymbol{r},t)$, followed by invoking the Biot-Savart law of magnetostatics, as follows:

$$\boldsymbol{B}(\boldsymbol{r},t) = \frac{\mu_0}{4\pi}\iiint_{-\infty}^{\infty} \boldsymbol{J}_{\text{bound}}(\boldsymbol{r}',t)\times\frac{\boldsymbol{r}-\boldsymbol{r}'}{|\boldsymbol{r}-\boldsymbol{r}'|^3}dx'dy'dz'. \qquad \text{(B1)}$$

Note that the Bio-Savart law ignores the time-dependence of $\boldsymbol{m}(t)$ by ignoring the effects of retardation. Integration over the volume of space between the parallel plates of Fig.3 then yields

$\boldsymbol{A}\times(\boldsymbol{B}\times\boldsymbol{C}) = (\boldsymbol{A}\cdot\boldsymbol{C})\boldsymbol{B} - (\boldsymbol{A}\cdot\boldsymbol{B})\boldsymbol{C}$

$$\boldsymbol{p}_{em}^{(\text{Livens})}(t) = \varepsilon_0\iiint_{-\infty}^{\infty}\boldsymbol{E}(\boldsymbol{r},t)\times\boldsymbol{B}(\boldsymbol{r},t)dxdydz$$
$$= \frac{\sigma_0}{4\pi}\iiint_{-\infty}^{\infty}\int_{x=-d/2}^{+d/2}\int_{y=-\infty}^{\infty}\int_{z=-\infty}^{\infty}\hat{\boldsymbol{x}}\times\left\{[\boldsymbol{\nabla}\times\boldsymbol{M}(\boldsymbol{r}',t)]\times\frac{\boldsymbol{r}-\boldsymbol{r}'}{|\boldsymbol{r}-\boldsymbol{r}'|^3}\right\}dxdydz\,dx'dy'dz'$$
$$= \frac{\sigma_0}{4\pi}\iiint_{-\infty}^{\infty}[\boldsymbol{\nabla}\times\boldsymbol{M}(\boldsymbol{r}',t)]\int_{x=-d/2}^{+d/2}\int_{y=-\infty}^{\infty}\int_{z=-\infty}^{\infty}\frac{x-x'}{|\boldsymbol{r}-\boldsymbol{r}'|^3}dxdydz\,dx'dy'dz'$$
$$-\frac{\sigma_0}{4\pi}\iiint_{-\infty}^{\infty}\hat{\boldsymbol{x}}\cdot[\boldsymbol{\nabla}\times\boldsymbol{M}(\boldsymbol{r}',t)]\int_{x=-d/2}^{+d/2}\int_{y=-\infty}^{\infty}\int_{z=-\infty}^{\infty}\frac{\boldsymbol{r}-\boldsymbol{r}'}{|\boldsymbol{r}-\boldsymbol{r}'|^3}dxdydz\,dx'dy'dz'$$
$$= \frac{\sigma_0}{2}\iiint_{-\infty}^{\infty}[\boldsymbol{\nabla}\times\boldsymbol{M}(\boldsymbol{r}',t)]\int_{x=-d/2}^{+d/2}\int_{\rho=0}^{\infty}\frac{(x-x')\rho}{[(x-x')^2+\rho^2]^{3/2}}dxd\rho\,dx'dy'dz'$$

$$-\frac{\sigma_0}{2}\iiint_{-\infty}^{\infty}[\boldsymbol{\nabla}\times\boldsymbol{M}(\boldsymbol{r}',t)]_x \int_{x=-d/2}^{+d/2}\int_{\rho=0}^{\infty}\frac{(x-x')\rho}{[(x-x')^2+\rho^2]^{3/2}}\,dx d\rho\, dx'dy'dz'$$

($x$ component of curl; Terms containing $\hat{\boldsymbol{y}}$ and $\hat{\boldsymbol{z}}$ vanish)

$$=\frac{\sigma_0}{2}\iiint_{-\infty}^{\infty}[\boldsymbol{\nabla}\times\boldsymbol{M}(\boldsymbol{r}',t)]_{y,z}\int_{x=-d/2}^{+d/2}\frac{(x-x')}{\sqrt{(x-x')^2}}\,dx\, dx'dy'dz'$$

($y,z$ components of curl)

$$=\frac{\sigma_0}{2}\iiint_{-\infty}^{\infty}[\boldsymbol{\nabla}\times\boldsymbol{M}(\boldsymbol{r}',t)]_{y,z}\big(|x-x'|_{x=-d/2}^{+d/2}\big)dx'dy'dz'$$

$$=-\sigma_0\iiint_{-\infty}^{\infty}\big[(\partial_{z'}M_x-\partial_{x'}M_z)\hat{\boldsymbol{y}}+(\partial_{x'}M_y-\partial_{y'}M_x)\hat{\boldsymbol{z}}\big]x'dx'dy'dz'$$

(Integration by parts)

$$=\sigma_0\iiint_{-\infty}^{\infty}x'\partial_{x'}\big(M_z\hat{\boldsymbol{y}}-M_y\hat{\boldsymbol{z}}\big)dx'dy'dz'=-\sigma_0\iiint_{-\infty}^{\infty}\big(M_z\hat{\boldsymbol{y}}-M_y\hat{\boldsymbol{z}}\big)dx'dy'dz'$$

$$=\sigma_0\hat{\boldsymbol{x}}\times\iiint_{-\infty}^{\infty}\boldsymbol{M}(\boldsymbol{r}',t)dx'dy'dz'=\varepsilon_0\boldsymbol{E}_0\times\boldsymbol{m}(t). \qquad \text{(B2)}$$

The above result is the same as that used in Eq.(53). While the Abraham momentum of the EM field in the system of Fig.3 is equal to zero, the Livens momentum has a contribution arising from the interaction between the external $E$-field and the magnetic dipole moment of the particle.